\documentclass[prx,amsmath,twocolumn,showpacs,superscriptaddress]{revtex4}

\usepackage[toc]{appendix}
\usepackage{epsfig}
\usepackage{amsfonts} 
\usepackage{color} 
\usepackage{hyperref}
\usepackage{natbib} 
\usepackage{graphicx,amssymb} 
\usepackage{multirow}
\newcommand{\WK}{Wiener-Khinchin theorem }

\begin{document}

\title{Aging Wiener-Khinchin Theorem and Critical Exponents of $1/f$ Noise }
\author{N. Leibovich}
\affiliation{Department of Physics, Institute of Nanotechnology and Advanced Materials,  Bar Ilan University, Ramat-Gan
52900, Israel}
\author{A. Dechant}
\affiliation{Department of Physics, Institute of Nanotechnology and Advanced Materials,  Bar Ilan University, Ramat-Gan
52900, Israel}
\affiliation{Department of Physics, Friedrich-Alexander-Universit\"at Erlangen-N\"urnberg, 91058 Erlangen, Germany}
\author{E. Lutz}
\affiliation{Department of Physics, Friedrich-Alexander-Universit\"at Erlangen-N\"urnberg, 91058 Erlangen, Germany}
\author{E. Barkai }
\affiliation{Department of Physics, Institute of Nanotechnology and Advanced Materials,  Bar Ilan University, Ramat-Gan
52900, Israel}

%%%%%%%%%%%%%%%%%%%%%%%%%%%%%%%%%%%%%%%%%%%%%%%%%%%%%%%%%%%%%%%%%%%%%%%%%%%%%%%
%
% A B S T R A C T 
%
%%%%%%%%%%%%%%%%%%%%%%%%%%%%%%%%%%%%%%%%%%%%%%%%%%%%%%%%%%%%%%%%%%%%%%%%%%%%%%%
\begin{abstract}
The power spectrum of a stationary process may be calculated in terms of the autocorrelation function using the Wiener-Khinchin theorem. We here generalize the Wiener-Khinchin theorem for nonstationary processes and introduce a time-dependent power spectrum $\left\langle S_{t_m}(\omega)\right\rangle$ where $t_m$ is the measurement time.  For processes with an aging correlation function of the form $\left\langle I(t)I(t+\tau)\right\rangle=t^{\Upsilon}\phi_{\rm EA}(\tau/t)$, where $\phi_{\rm EA}(x)$ is a nonanalytic function when $x$ is small, we find aging $1/f$ noise. Aging $1/f$ noise is characterized by five critical exponents.
We derive the relations between the scaled correlation function and these exponents. We show that our definition of the time-dependent spectrum retains its interpretation as a density of Fourier modes and discuss the relation to the apparent infrared divergence of $1/f$ noise. We illustrate our results for blinking quantum dot models, single-file diffusion and Brownian motion in logarithmic potential.  
\end{abstract}


\maketitle

\section{Introduction}
In many applications, a random process $I(t)$ recorded in time interval $[0,t_m]$, is analyzed using the sample spectrum,  $S_{t_m}(\omega)=|\int_0^{t_m}I(t')\exp(-\imath \omega t'){\rm d}t'|^2/t_m$, where the measurement time $t_m$ is assumed to be long.  
For a stationary process, the power spectrum is routinely calculated from the autocorrelation function, $C(\tau)=\left\langle I(t)I(t+\tau)\right\rangle$, using the Wiener-Khinchin theorem \cite{Kubo,Priestley};
\begin{equation}
\lim_{t_m\rightarrow\infty}\left\langle S_{t_m}(\omega)\right\rangle=2\int_{0}^\infty{\rm d}\tau C(\tau)\cos(\tau \omega).
\label{eq:WK}
\end{equation}
Obviously not all physical processes are stationary \cite{Bouchaud,Aquino,Eliazar,Rod,Bouchaud96,Silv,Niemann,Schriefl,Cugliandolo,LFC,Bellon,Crisanti} and then the \WK \eqref{eq:WK} does not hold. Extending the \WK to non-stationary processes has been a topic of many works \cite{Cohen,Page,Lukovic,Jung}. Some relate it to the instantaneous power spectrum, where it looses its meaning as a density since it may get negative values (e.g. \cite{Cohen,Page}). Others deal with a specific process such as telegraphic noise \cite{Lukovic} or periodically driven stochastic systems \cite{Jung}. % Here we address a proper (positive) time-dependent spectral density, for general nonstationary processes.
%However, the scope of the nonstationarity in general is too wide for such a generalization to be meaningful. Luckily, in many systems studied over the last decades, a particular type of aging correlation function, $C(t,\tau)\sim t^\gamma\phi_{\rm EA}(\tau/t)$ emerges, where $(.)_{\rm EA}$ refers to ensemble average. 

We here consider the power spectrum of systems exhibiting scale invariant aging %. Here the process is stationary in the mean sense, namely $\langle I \rangle$ is time independent. We investigate processes 
with a correlation function of the form $C(t,\tau)\sim t^\Upsilon\phi_{\rm EA}(\tau/t)$, where $(.)_{\rm EA}$ refers to ensemble average. 
Such correlation functions appear in a vast array of systems and models ranging from glassy dynamics \cite{Bouchaud,Dean,Bertin}, blinking quantum dots \cite{Margolin04}, laser-cooled atoms \cite{Dechant2012,Kessler}, motion of a tracer particle in a crowded environment \cite{Lizana,Leibovich}, elastic models of fluctuating interfaces \cite{Taloni}, diffusion in heterogeneous environment \cite{Chevstry}, deterministic noisy Kuramoto models \cite{Ionita}, granular gases \cite{Bordova}, deterministic intermittency  \cite{AgingELI}, to growing interfaces following the KPZ equation \cite{TakeuchiJSP}. %, to name only some examples. In some cases the scaling function exhibits a second scaling exponent, $\langle I(t) I(t+\tau) \rangle\sim t^\Upsilon \phi_{{\rm EA}}(\tau/t^\zeta)$,  or even a logarithmic time dependence \cite{Bertin}.  Here we will avoid this zoo of exponents and focus on the classification of the spectrum for the case $\zeta=1$. 
%In our recent publications \cite{DechantPRL,LeibovichPRL} we have briefly introduced the relation between the sample spectrum and the aging correlation function, thus generalizing the \WK to aging systems. Here we present a more detailed derivation of the relation between the sample spectrum and the nonstationary correlation function in Secs. II and III. This relation provides a direct extension of the \WK to this broad class of aging dynamics.
In our recent publications \cite{DechantPRL,LeibovichPRL}, we have generalized the \WK to these aging processes by introducing a time-dependent spectral density. We have moreover established a correspondence to $1/f$ noise when $\phi_{\rm EA}(x)$ is not analytic for small argument.   

The power spectrum of $1/f$ noise at low frequencies is: 
\begin{equation}
S(\omega)\sim \omega^{-\beta} \ \ \ \ 0<\beta<2.
\label{GeneralPS.eq}
\end{equation}
The value $\beta=0$ corresponds to white noise and $\beta=2$ to Brownian noise. 
$1/f$ noise, with a range of different exponents, occurs in many systems in a variety of disciplines. A partial list includes electronic, solid and condensed matter devices \cite{Dutta, Keshner, Hooge81, Weissman}, sand-pile models \cite{Yadav}, blinking quantum dots \cite{Frantsuzov,Pelton2004}, nanoscale electrodes \cite{Kraft}, experimental data of voltage-dependent anion channel in rats brains \cite{Banerjee} and processes modeled by nonlinear-stochastic-differential equations \cite{Kaulakys}. 

In \cite{Mandelbrot1967} Mandelbrot suggested that ``one needs a non-Wienerian spectral theory to account for $f^{\theta-2}$ noise'', where a time-series-length-dependent spectrum might be measured \cite{Mandelbrot1967,Graves}. Indeed, as shown theoretically \cite{Margolin06, Niemann,Rod} and experimentally \cite{Kraft,Sadegh}, the power spectrum of a blinking quantum dot and of nano electrodes ages, namely as the measurement time becomes longer the intensity of the measured noise is reduced, decaying as a power law.  

\begin{figure}
	\centering
		\includegraphics[width=\columnwidth]{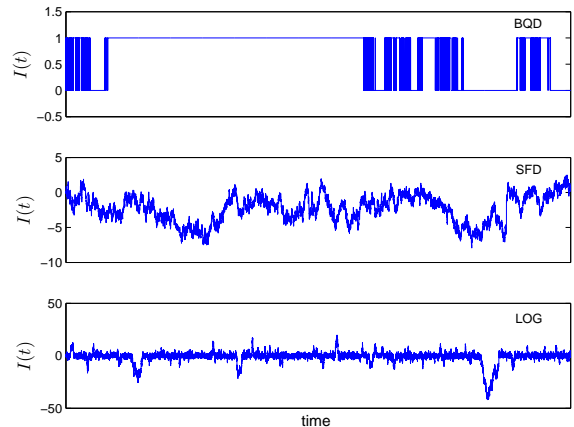}
		\caption{The time trace of three models: blinking quantum model with $\alpha=0.8$ (upper), single file diffusion with $D=0.5$ and $a=1$ (middle), and Brownian motion in a logarithmic potential (lower panel). See details on each model in Sects. \ref{BlinkingQD},\ref{SFD} and  \ref{LogPotential} (respectively). }
	\label{fig:trajectories}
\end{figure}

Traditional studies of $1/f^\beta$ noise characterize the spectrum with a single exponent $\beta$.  In the recent experiment \cite{Sadegh}, the aging properties of $1/f$ noise were characterized with the help of five different exponents, $\beta,z,\mu,\eta$ and $\delta$ defined as follows: the asymptotic power spectrum is of the form $S(\omega)\sim A_{t_m}\omega^{-\beta}$ with the time-dependent amplitude $A_{t_m}\sim (t_m)^{-z}$ for long times and low-frequencies cutoff $\omega_{\rm min}\sim (t_m)^{-\eta}$. Furthermore, the ``power'' at zero frequency is $S(0)\sim (t_m)^{\mu}$ and the total measured power $\int_{1/t_m}^{\infty}S(\omega){\rm d}\omega\sim (t_m)^{\delta}$. Our aim is here to derive these five exponents from the correlation function for general aging processes and investigate their relationships. We apply our finding to blinking quantum dot  models \cite{Margolin04}, as well as to single file diffusion \cite{Leibovich,Lizana} and diffusion in a logarithmic potential \cite{Kessler,Dechant2012}. Sample paths of these models, i.e. a representative time trace of these processes, are shown in Fig. \ref{fig:trajectories}. Visually these processes appear very different, the underlying unifying theme is their description in terms of a scale invariant  correlation function which leads to $1/f$ noise. 

For stationary processes, the spectrum is related to the discrete Fourier modes at the natural frequencies $\omega_n = 2 \pi n/t_m$ with $n$ integer \cite{Kubo}. The integral of the spectrum over all frequencies is finite and equal to the sum of the Fourier modes' intensities. Since the spectrum is thus positive and normalizable, we can interpret it as the continuous distribution of Fourier modes in the limit of infinite measurement time. By contrast, $1/f$ noise with $\beta > 1$ exhibits a non-integrable infrared divergence and thus infinite power, which has led to some discussions on its physical interpretation \cite{Niemann,Mandelbrot1967,Graves}. We show that allowing for a measurement-time dependent spectrum naturally resolves this apparent paradox, as the power remains finite at finite times. Thus even for nonstationary processes, the spectrum retains its interpretation as a density in frequency space. While for finite measurement times, the detectable Fourier modes are of course discrete, we argue that nevertheless, the spectrum can also be understood as a continuous function of frequency. This function then exhibits scale invariant oscillations, which can yield additional information on the critical exponents not contained in the natural frequencies.

The outline of the paper is the following. In Sects. \ref{sec2} and \ref{sec3}, we derive the generalized \WK connecting the aging correlation function $C(t,\tau)$ and the time-dependent spectral density $S_{t_m}(\omega)$. In Sect. \ref{1/f noise} we evaluate explicitly the spectral density for a scale invariant correlation function and discuss the relationship with $1/f$ noise. We then compute in Sect. \ref{CriticalExponents} the five critical exponents characterizing $1/f$ noise. In Sect.~\ref{Sec6}, we compare the properties of the time dependent spectrum to stationary case and discuss its relation to Fourier modes. Finally, in Sects. \ref{BlinkingQD}, \ref{SFD} and \ref{LogPotential}, we apply our results to concrete examples: blinking quantum dot, single file diffusion and diffusion in a logarithmic potential.   

\section{Aging Power Spectrum: The Time-Averaged Formalism}
\label{sec2}
For  a nonstationary process, the ensemble-average autocorrelation function is  $\left\langle I(t)I(t+\tau)\right\rangle=C(t,\tau)$ and depends explicitly on the time $t$ and the lag time $\tau$.
To examine analytically the temporal behavior of the power spectrum, we follow Mandelbrot \cite{Mandelbrot1967} and define the power spectrum of a random signal $I(t)$ as $S_{t_m}(\omega)=|\hat{I}_{t_m}(\omega)|^2/{t_m}$ where $\hat{I}_{t_m}(\omega)=\int_0^{t_m}I(t')\exp(-\imath\omega t'){\rm d}t'$, and $t_m$ is the measurement time.
The spectrum, by this definition, is, 
\begin{equation}
\left\langle S_{t_m}(\omega)\right\rangle=\left\langle \frac{1}{t_m} \int_0^{t_m}{\rm d}t_1I(t_1)e^{-\imath\omega t_1}\int_0^{t_m}{\rm d}t_2I(t_2)e^{+\imath\omega t_2}\right\rangle.
\end{equation}
Equivalently we can write,
\begin{eqnarray}
\left\langle S_{t_m}(\omega)\right\rangle =&& \frac{1}{t_m}\int_0^{t_m}{\rm d}t_1\int_{t_1}^{t_m}{\rm d}t_2\left\langle I(t_1)I(t_2)\right\rangle e^{-\imath\omega(t_1-t_2)} \nonumber\\ 
+&& \frac{1}{t_m}\int_0^{t_m}{\rm d}t_1\int_0^{t_1}{\rm d}t_2\left\langle I(t_1)I(t_2)\right\rangle e^{-\imath\omega(t_1-t_2)} , \nonumber \\
\label{eq:4}
\end{eqnarray}
where the first term on the right side corresponds to $t_1<t_2$, and the second term is for $t_2<t_1$. In order to express Eq.~\eqref{eq:4} in terms of the correlation function we first change the integration order and the names of integration variables ($t_1 \leftrightarrow t_2$) to find
\begin{eqnarray}
&&\left\langle S_{t_m}(\omega)\right\rangle =  \\&& \frac{1}{t_m}\int_0^{t_m}{\rm d}t_1\int_{t_1}^{t_m}{\rm d}t_2\left\langle I(t_1)I(t_2)\right\rangle \left[e^{\imath\omega(t_1-t_2)}+e^{-\imath\omega(t_1-t_2)}\right]. \nonumber
\end{eqnarray} 
An additional change of variables, $t_1=t_1$ and $t_2-t_1=\tau$, gives
\begin{equation}
\left\langle S_{t_m}(\omega)\right\rangle=\frac{2}{t_m}\int_0^{t_m}{\rm d}t_1\int_0^{t_m-t_1} d\tau \left\langle I(t_1)I(t_1+\tau)\right\rangle \cos(\omega\tau).
\end{equation}
By interchanging the order of integration, we finally obtain \cite{Margolin06}
\begin{equation}
\left\langle S_{t_m}(\omega)\right\rangle=\frac{2}{t_m}\int_0^{t_m}{\rm d}\tau (t_m-\tau)\left\langle C_{\rm TA}(t_m,\tau)\right\rangle \cos(\omega\tau),
\label{eq:GenerlizedWK1}
\end{equation}
where the time-averaged correlation function is defined as 
\begin{equation}
C_{\rm TA}(t_m,\tau)=\frac{1}{t_m-\tau}\int_0^{t_m-\tau}{\rm d}t_1I(t_1)I(t_1+\tau).
\label{eq:C3}
\end{equation}
We emphasize that the ensemble- and time-average correlation functions, $C(t,\tau)$ and $C_{\rm TA}(t_m,\tau)$, are not identical for the underlying processes considered in this paper. %We recall the physical examples mentioned in the introduction Refs. \cite{Bouchaud,Dean,Bertin,Margolin04,DechantPRX13,Lizana,Leibovich,Taloni,Ionita,Bordova,AgingELI,TakeuchiJSP} and
In the following, we consider time-averaged correlation functions of the scaling form 
\begin{equation}
\left\langle C_{TA}(t_m,\tau)\right\rangle=t_m^{\Upsilon}\varphi_{\rm TA}(\tau/t_m).
\label{eq:Eq9}
\end{equation} 
For the time being, we assume that  this asymptotic scaling is valid for all $\tau$ and $t_m$. For real physical systems this is an approximation, as we discuss in Sects.  \ref{BlinkingQD}-\ref{LogPotential}. We also assume $\langle I \rangle$ is a constant independent of time, see further discussion in Appendix \ref{sec:bounds}.
Substituting Eq.~\eqref{eq:Eq9} in Eq.~(\ref{eq:GenerlizedWK1}) and changing variables ($x=\tau/t_1$), we obtain what we call the time-averaged form of the aging Wiener-Khinchin theorem,
\begin{equation}
\left\langle S_{t_m}(\omega)\right\rangle=2t_m^{\Upsilon+1}\int_0^1{\rm d}x(1-x)\varphi_{\rm TA}(x) \cos(\omega t_m x).
\label{eq:GenerlizedWK}
\end{equation}
By using a cosine transform \cite{Boas} we further find 
\begin{equation}
t_m^{\Upsilon}\varphi_{\rm TA}(x)=\frac{1}{2\pi(1-x)}\int_{-\infty}^{\infty}{\rm d}\omega\left\langle S_{t_m}(\omega)\right\rangle \cos(\omega t_m x),
\label{eq:Inverse}
\end{equation}
where $0<x<1$ and $\left\langle S_{t_m}(\omega)\right\rangle$ is an even function.
%In what follows we obtain the aging power spectrum from the underlying correlations functions of physical systems. In Appendix \ref{sec:inverse} we discuss briefly the inverse operation, namely given the power spectrum we estimate the aging correlation function for a model blinking quantum dot. 
Equation \eqref{eq:Inverse} established a direct connection between the scaling function $\varphi_{\rm TA}(x)$ and the average power spectrum $\left\langle S_{t_m}(\omega)\right\rangle$. Care must be taken when evaluating information on the underlying correlation function from the sample spectrum, see discussion in Appendix \ref{sec:inverse}.

\section{Aging Power Spectrum: The Ensemble-Averaged Formalism}
\label{sec3}
In the present section we derive %Here we present
 a relation between the ensemble-averaged autocorrelation function  $C(t,\tau)$ and the average time-dependent power spectrum $\left\langle S_{t_m}(\omega)\right\rangle$. We assume, as before, that the correlation function scales as 
\begin{equation}
C(t,\tau)=t^{\Upsilon}\phi_{\rm EA}(\tau/t).
\label{eq:ScaledCorr}
\end{equation}
at all times $\tau$ and $t$. Systems which exhibit this type of correlation scaling behavior have been discussed in Refs. \cite{Bouchaud,Dean,Bertin,Margolin04,DechantPRX13,Lizana,Leibovich,Taloni,Ionita,Bordova,AgingELI,TakeuchiJSP}. By taking the ensemble average of Eq.~\eqref{eq:C3}, we directly obtain a connection between time- and ensemble- averaged scaling function
%By averaging upon Eq. \eqref{eq:C3} we find the relation between the scaling of the ensemble-average correlation function and time-average correlation function;
\begin{equation}
\varphi_{\rm TA}(x)=\frac{x^{\Upsilon+1}}{1-x}\int_{\frac{x}{1-x}}^{\infty}{\rm d}y\frac{\phi_{\rm EA}(y)}{y^{2+\Upsilon}}.
\label{eq:CorrTA.EN}
\end{equation}
%The ensemble-average correlation function in terms of the time-average correlation function is given by
%\begin{eqnarray}
%&&\phi_{EA}(y)=\varphi_{TA}\left(\frac{y}{1+y}\right)\left[(\Upsilon+1)(1+y)^{\Upsilon}-\Upsilon y(1+y)^{\Upsilon-1}\right]\nonumber\\
%&&-\frac{d}{dy}\varphi_{TA}\left(\frac{y}{1+y}\right)\cdot y(1+y)^{\Upsilon}.
%\end{eqnarray} 

%To start with, and for the sake of simplicity, we 
Let us first assume $\Upsilon=0$. Then substituting Eq.~\eqref{eq:CorrTA.EN} in Eq.~(\ref{eq:GenerlizedWK}) and changing the integration order we find
\begin{equation}
\left\langle S_{t_m}(\omega)\right\rangle=2t_m\int_0^{\infty}{\rm d}x\phi_{\rm EA}(x)\int_0^{1/(1+x)}{\rm d}\tilde{t_1}\cos(\tilde{\omega} \tilde{t_1}x)\tilde{t_1}.
\label{eq:gamma0}
\end{equation}
Integrating by parts gives 
\begin{eqnarray}
&&\left\langle S_{t_m}(\omega)\right\rangle= \label{eq:12}\\ \nonumber
&&2{t_m}\int_0^{\infty}{\rm d}x\phi_{\rm EA}\left(x\right)\left[\frac{\sin(\tilde{\omega}\frac{x}{1+x})}{\tilde{\omega x}(1+x)}+ 
\frac{\cos(\tilde{\omega}\frac{x}{1+x})}{\tilde{\omega}^2x^2}-\frac{1}{\tilde{\omega}^2x^2} \right] .
\end{eqnarray}
Again changing variables  according to $y=x/(1+x)$ eventually gives 
\begin{eqnarray}
&&\left\langle S_{t_m}(\omega)\right\rangle= \label{eq:InfiniteCovarianceDensity} \\ \nonumber
&&2t_m\int_0^1\frac{{\rm d}y}{(\tilde{\omega}y)^2}\phi_{\rm EA}\left(\frac{y}{1-y}\right)\left[\tilde{\omega}y\sin(\tilde{\omega}y)+\cos(\tilde{\omega}y)-1\right]
\end{eqnarray}
where $\tilde{\omega}=\omega t_m$.

For $\Upsilon\neq 0$ we find in a similar manner 
\begin{equation}
\left\langle S_{t_m}(\omega)\right\rangle=2t_m\int_0^{\infty}{\rm d}x\phi_{\rm EA}(x)\int_0^{1/(1+x)}{\rm d}\tilde{t_1}\cos(\tilde{\omega} \tilde{t_1}x)\tilde{t_1}^{\Upsilon+1}.
\label{eq:gamma}
\end{equation}
%instead of Eq. (\ref{eq:gamma0}). Hence we find using \cite{Abramowitz}
Evaluating the last integral explicitly \cite {Abramowitz}, we finally arrive at
\begin{widetext}
\begin{equation}
\left\langle S_{t_m}(\omega)\right\rangle=
\frac{2t_m^{\Upsilon+1}}{2+\Upsilon}\int_0^1{\rm d}y(1-y)^{\Upsilon}\phi_{\rm EA}\left(\frac{y}{1-y}\right)  {_1F_2}\left[1+\frac{\Upsilon}{2};\frac{1}{2},2+\frac{\Upsilon}{2};-\left(\frac{\tilde{\omega}y}{2}\right)^2\right], 
\label{eq:GeneralizedInfiniteCovarianceDensity} 
\end{equation}
\end{widetext}
%The integral convergence demands that 
%\begin{eqnarray}
%\lim_{y\rightarrow 1}\phi_{\rm EN}(\frac{y}{1-y})<(1-y)^{m_1} \ \ \ \ && m_1>-\Upsilon-1,   \\ \nonumber
%\lim_{y\rightarrow 0}\phi_{\rm EN}(\frac{y}{1-y})<y^{m_2} \ \ \ \ &&m_2>-1.
%\end{eqnarray}
where $_1F_2[a;b_1,b_2;z]$ refers to the hypergeometric function and $\Upsilon>-2$ for convergence. Equation \eqref{eq:GeneralizedInfiniteCovarianceDensity} is our second aging \WK connecting the ensemble-average scaling correlation function $\phi_{\rm EA}$ to the sample spectrum.   
The inverse formula, which relates the power spectrum to the ensemble-average correlation function, is
\begin{eqnarray}
&&C(t_m-\tau,\tau)= \label{eq:Inv} \\ 
&&\frac{1}{\pi}\int_0^{\infty}{\rm d}\omega\cos(\omega\tau)\left[\left\langle S_{t_m}(\omega)\right\rangle+t_m\frac{\partial}{\partial t_m}\left\langle S_{t_m}(\omega)\right\rangle\right] .
\nonumber
\end{eqnarray}
This inversion is general and valid for any type of correlation function.

%Here, following \cite{LeibovichPRL,DechantPRL}, we have formulated two aging Wiener-Khinchin theorems,  Eqs. (\ref{eq:GenerlizedWK},{\ref{eq:GeneralizedInfiniteCovarianceDensity}), relating between time- and ensemble- average correlation functions and the sample spectrum, where the choice of theorem to be used in practice depends on the application. Most theoretical works provide an ensemble average correlation function.
Equations (\ref{eq:GenerlizedWK}) and ({\ref{eq:GeneralizedInfiniteCovarianceDensity}) provide two forms of aging Wiener-Khinchin theorem, relating the sample spectrum to either the time- or ensemble- average scaling correlation function. 
The choice between the theorems depends on the practical application. Most theoretical works provide an ensemble average correlation function $\phi_{\rm EA}(x)$.
In this case,  to use Eq.~(\ref{eq:GenerlizedWK})  we need to determine 
the time averaged correlation function 
from Eq.~\eqref{eq:CorrTA.EN} first.
On the other hand, to use Eq.~\eqref{eq:GeneralizedInfiniteCovarianceDensity} we need to determine the time dependency of the correlation function, in particular the exponent $\Upsilon$, which in experimental situations is a-priori unknown, though it could be estimated from data. In addition, the inverse formula Eq.~\eqref{eq:Inv} contains a derivative term, which may increase measurement errors (see Appendix \ref{sec:inverse} where the inversion is performed for a specific example). Still, both formalisms are clearly equivalent and useful. 

\section{Scale-Invariant Correlation Function and the Power Spectrum}
\label{1/f noise}
We next compute the power spectrum for a scaling function of the form 
\begin{equation}
\phi_{\rm EA}(x) \approx   \left\{
\begin{array}{llll}
a_0-a_{{\rm V}}x^{{\rm V}} & & & x\ll 1\\
b_0-b_{\Lambda}x^{\Lambda} & & & x \gg 1
\end{array}
		 \right.
		\label{eq:ScaledCorr1}.
\end{equation}
where $a_0$, $b_0$, $a_{{\rm V}}$ and $b_{\Lambda}$ are constants which are determined by the specific process. %Notice that generally this approximation may contain more then two component with exponents less then 1, but higher order then $N$ does not contribute to the spectrum when $\tilde{\omega}\gg 1$, therefore we neglect them. 
Using Eq.~(\ref{eq:CorrTA.EN}) we find the time averaged correlation function
\begin{equation}
\varphi_{\rm TA}(x) \approx \left \{  
\begin{array}{lll}
\tilde{a}_0-\tilde{a}_{{\rm V}}x^{{\rm V}} & & x\ll 1\\
\tilde{b}_0(1-x)^{\Upsilon}-\tilde{b}_{\Lambda}(1-x)^{\Upsilon-\Lambda} & & 1-x \ll 1
\end{array}
		 \right.
\end{equation}
where $\tilde{a}_0=a_0/(1+\Upsilon)$, $\tilde{b}_0=b_0/(1+\Upsilon)$, $\tilde{a}_{{\rm V}}=a_{{\rm V}}/(1+\Upsilon-{\rm V})$ and $\tilde{b}_{\Lambda}=b_{\Lambda}/(1+\Upsilon-\Lambda)$.  We assume $0<|{\rm V}|<1$, $\Upsilon-{\rm V}>-1$, $\Lambda<0$ and $\Upsilon-\Lambda>-1$  for convergence.
These conditions are naturally satisfied for all relevant examples, see Table.~\ref{Tab1} and Sects. \ref{BlinkingQD}-\ref{LogPotential}.

Using Eq.~(\ref{eq:GenerlizedWK}), the power spectrum for such a process in the limit $\omega t_m =2\pi n$ where $n$ is a large positive integer, is
\begin{equation}
\left\langle S_{t_m}(\omega)\right\rangle_{\omega t_m =2\pi n}\approx 2\tilde{a}_{{\rm V}}\frac{\sin(\pi {\rm V}/2)\Gamma(1+{\rm V})}{t_m^{-\Upsilon+{\rm V}}\omega^{1+{\rm V}}}.
\label{eq:1FNoise}
\end{equation}
Accordingly, the scale-invariant correlation function \eqref{eq:ScaledCorr1} leads to $1/f$ noise.  
The next leading terms are 
\begin{eqnarray}
&&\left\langle S_{t_m}(\omega)\right\rangle_{\omega t_m \gg 1}\approx 2\tilde{a}_{\rm V}\frac{\sin(\pi {\rm V}/2)\Gamma(1+{\rm V})}{t_m^{-\Upsilon+{\rm V}}\omega^{1+{\rm V}}}     \nonumber\\
&&+\frac{2\tilde{a}_0}{\omega^2t_m^{-\Upsilon+1}}-2\tilde{b}_0\frac{\Gamma(2+\Upsilon)\cos(\omega t_m-\Upsilon\pi/2)}{\omega^{2+
\Upsilon}t_m} \nonumber\\
&&+2\tilde{b}_{\Lambda}\frac{\Gamma(2+\Upsilon-\Lambda)\cos\left[\omega t_m-(\Upsilon-\Lambda)\pi/2\right]}{\omega^{2+
\Upsilon-\Lambda}t_m^{-\Lambda+1}}. \label{eq:NonStationatyPS}
\end{eqnarray}
When $\omega t_m$ is treated as a continuous variable, Eq.~\eqref{eq:NonStationatyPS} exhibits oscillations. For specific examples, these oscillations are discussed in Sects. \ref{BlinkingQD}-\ref{LogPotential}. 
The conditions on the exponents $\Upsilon,{\rm V}$ and $\Lambda$ guarantee that the $1/f$ spectrum Eq.~\ref{eq:1FNoise} is indeed the leading order for large $\omega t_m$. Exact $1/f$ noise, with $\beta=1$ and logarithmic time dependence \cite{Rod}, is not discussed here and left for a future work.  
In Sects. \ref{BlinkingQD}-\ref{LogPotential} we show that this result is not valid for arbitrarily large frequencies, since then the scaling assumption breaks down.

%From Eq. \eqref{eq:NonStationatyPS} we conclude that for a correlation function that scales as Eq. \eqref{eq:ScaledCorr1}, we expect $1/f$ noise with oscillations where their amplitude decreasing with frequency. 
We note that when $\omega t_m\gg 1$ the spectrum is controlled by the first term, which was determined by the nonanalytic expansion of the correlation function when $\tau\ll t_m$. The oscillating behavior seen in Eq.~\eqref{eq:NonStationatyPS} is a finite-measurement-time effect and is related to the correlation function when  $\tau\sim t_m$. Thus detecting these oscillations gives insight on the details of the underlying correlation function. In fact since $1/f$ spectrum is so common, yet its physical origin still remains unclear in many cases, the oscillating part of the spectrum might be a valuable tool for distinguishing between microscopic models.
These oscillations depend only on the scaling variable $\omega t_m$ and are universal in that sense. In our examples below Eq.~\eqref{eq:1FNoise} works well also when $\omega t_m \approx 1$.% especially when the power spectrum is plotted on log-log scale deviations from the limit Eq.~\eqref{eq:NonStationatyPS} might be difficult to detect.

We note that when the spectrum is evaluated on the natural frequencies $\omega t_m=2\pi n$, $n\in{\mathbb N}$, then according to Eq. \eqref{eq:1FNoise} the power spectrum is characterized by two exponents; $\Upsilon$ and ${\rm V}$. These are given in Table~\ref{Tab1} for specific systems exhibiting aging. When we consider the continuous frequencies then, according to Eq.~\eqref{eq:NonStationatyPS}, the spectrum is quantified using three exponents $\Upsilon$, ${\rm V}$ and $\Lambda$ describing the time and frequency dependence of the $1/f$ spectrum. Below we discuss three additional exponents, which characterize the process.

\begin{table*}	
\center{
	\begin{tabular}{|c|c|c|c|c|}  \hline
	   Model [Ref.] & &$\Upsilon$ &  ${\rm V}$ & \\ \hline
	 Unilayer Parisi's Tree \cite{Dean} & $0<\alpha<1$ & $0$   & $\alpha-1$ & analytic \\ \hline
	 Blinking Quantum Dot \cite{Margolin04} & $0<\alpha<1$ & $0$
	\footnote{Power-law `on'/`off' waiting time \cite{Margolin04}} \ ,\ \ $2\alpha-2$\footnote{finite mean `on' time \cite{Margolin04}} & $\alpha-1$\footnotemark[1] \ ,\ \ $1-\alpha$ \footnotemark[2] & analytic \\ \hline
	 Laser-Cooled Atoms \cite{Dechant2012} & $1<\alpha<3$&$2-\alpha$ & $2-\alpha$ & analytic \\ \hline
	 Single-File Diffusion \cite{Leibovich,Lizana} & & $1/2$ & $1/2$& analytic \\ \hline 
	 Generalized Elastic Model \cite{Taloni}& $0<\alpha<1$ & $\alpha$ & $\alpha$ & analytic \\ \hline
	 Coupled Classical Oscillators\footnote{Lattice size $32\times 32$, coupling strength $\kappa=-4$ \cite{Ionita}.}  \cite{Ionita} & &\ $[-0.02,0]$\footnote{Intermediate regime (ii) (see details in \cite{Ionita})}  \ \ \ $[-0.58,-0.4]$\footnote{Saturation regime (iii) (see details in \cite{Ionita})} \ & $-0.14\pm 0.03$\footnotemark[4] \ \ \ $0$\footnotemark[5] & numeric \\ \hline
	 1D Growing Interfaces  \cite{TakeuchiJSP} & & $\approx 0.33$ &  & experiment  \\ \hline 
	 Infinite RC transmission Line \cite{Keshner} &$1<\alpha<2$& $\alpha-1$ & $\alpha-1$ & analytic \\ \hline 
		\end{tabular}
		\caption{The aging behavior of several models, where the correlation function is given in terms of $\langle I(t)I(t+\tau)\rangle\sim t^{\Upsilon}\phi_{EA}(\tau/t)$ and $\phi_{EA}(x)\propto a_0- a_{{\rm V}}x^{{\rm V}}$ when $x\ll 1$. }
		\label{Tab1}}
\end{table*}
%\end{widetext}

\section{Critical Exponents and Scaling Relations}
\label{CriticalExponents}
As mentioned in the introduction, traditional theories of $1/f^\beta$ noise characterize the spectrum with a single exponent $\beta$. However, this is not sufficient as recent studies show \cite{Niemann,Rod,Kraft,Sadegh}.  
We follow \cite{Sadegh} and characterize the finite-time power spectra  $S_{t_m}(\omega)$ with five power laws as follows: (i) The power spectrum frequency dependence; $S(\omega)\sim \omega^{-\beta}$ for low frequencies, (ii) the power spectrum time dependence; $S(\omega)\sim A_{t_m}\omega^{-\beta}$ where $A_{t_m}\sim t_m^{-z}$ for long times, (iii) the lower cutoff time dependence; $\omega_{\rm min}\sim t_m^{-\eta}$, (iv) the power at zero frequency; $S(0)\sim t_m^{\mu}$ and (v) the total measured power; $\int_{1/t_m}^{\infty}S(\omega){\rm d}\omega\sim t_m^{\delta}$. In the present section we compute these exponents from the properties of the correlation function. We will later consider three physical models where these exponents are calculated explicitly; see Tables~\ref{Tab2} and \ref{Tab3} for a summary. 

%\subsection{Power Spectrum Frequency and Time Dependence and the Exponents $\beta$ and $z$}
We consider a process with a correlation function with the scaling behavior Eq.~(\ref{eq:ScaledCorr1}).
The critical exponent $\beta$ which is determined by the power-law decay of the average power spectra in Eq.~(\ref{eq:1FNoise}) is 
\begin{equation}
\beta=1+{\rm V},
\end{equation} 
where $\beta\neq 1$. In addition the aging exponent $z$ which is related to the time decay is 
\begin{equation}
z={\rm V}-\Upsilon,
\end{equation}
so clearly $\beta=1+ z + \Upsilon$.  When $\beta=1$ logarithmic time corrections are expected \cite{Rod}. In an experimental situation the exponent $\Upsilon$ may be measured through the zero-frequency power as described in the next paragraph.

\subsection{Zero-Frequency Power Density and the Exponent $\mu$}

To determine the exponent $\Upsilon$ one may measure the spectrum at zero frequency, i.e. $\left\langle S_{t_m}(\omega=0)\right\rangle=t_m\left\langle\overline{I}_{t_m}^2\right\rangle$ where the time-averaged signal is
$\overline{I}_{t_m}=\int_0^{t_m}I(t){\rm d}t/t_m$. 
Of course the zero frequency cannot be considered part of the spectrum itself, at least not
in the traditional sense, since in a finite time measurement one cannot detect a frequency shorter than $2\pi/t_m$. However, this does not imply that it cannot be measured, it is rather easy to do so. 
For a stationary process $\left\langle S_{t_m}(0)\right\rangle$ is linearly dependent on the measurement time $t_m$. For a nonstationary process its time dependence is related to the exponent $\Upsilon$ as follows.

When the ensemble-average correlation function is scaled as Eq.~\eqref{eq:ScaledCorr} the power density at zero frequency is 
\begin{equation}
\left\langle S_{t_m}(\omega=0)\right\rangle=2t_m^{\Upsilon+1}\int_0^{\infty}{\rm d}x\frac{\phi_{\rm EA}(x)}{(1+x)^{\Upsilon+2}},
\label{eq:S0.1}
\end{equation}
or equivalently (after a change of variables)
\begin{equation}
\left\langle S_{t_m}(\omega=0)\right\rangle=2t_m^{\Upsilon+1}\int_0^1{\rm d}x\phi_{\rm EA}\left(\frac{x}{1-x}\right).
\label{eq:ZeroFrequency}
\end{equation} 
The exponent $\mu$ is hence given by
\begin{equation}
\mu=1+\Upsilon.
\end{equation}
Notice that the scaling relation $\mu=\beta-z$ is also valid. This relation was suggested in Ref. \cite{Sadegh} in the context of blinking quantum dot models. 

{\bf Remark:} Equation (\ref{eq:S0.1}) was already obtained in the context of a scaling Green-Kubo relation \cite{DechantPRX13}. The scaling Green-Kubo formula expresses the relation between the diffusion coefficient %$D_\Upsilon$
of an enhanced diffusion process, %, where $\left\langle x^2(t)\right\rangle=2D_{\Upsilon}t^{\Upsilon+2}$,
and the scale-invariant velocity correlation function. % $C_v(t+\tau,t)={\cal{C}}t^{\Upsilon}\phi(\tau/t)$ where $\Upsilon>-1$. 
The mean square displacement is %defined by $\left\langle x^2(t)\right\rangle\approx2{\cal{C}}\int_0^{t}{\rm d}t_2\int_0^{t_2}{\rm d}t_1C_N(t_2,t_1)$ and it is 
equivalent to $t_m\left\langle S(0)\right\rangle$ which is the spectrum in zero-frequency multiplied by the measurement time $t_m$. 

\subsection{Lower Cutoff Time Dependence}
The lower cutoff time dependence $\omega_{\rm min}\sim t_m^{-\eta}$ is defined by the transition frequency between the power-law decay $\left\langle S_{t_m}(\omega)\right\rangle_{\omega t_m \gg 1}$  and $\left\langle S_{t_m}(0)\right\rangle$. By comparing Eq.~\eqref{eq:1FNoise} and Eq.~(\ref{eq:ZeroFrequency}), i.e. $\langle S_{t_m}(0)\rangle = A_{{\rm V}}t_m^{\Upsilon-{\rm V}}\omega_{\rm min}^{-1-{\rm V}}$ where $A_{{\rm V}}=2\tilde{a_{\rm V}}\sin(\pi {\rm V}/2)\Upsilon(1+{\rm V})$, we find 
\begin{equation}
\omega_{\rm min}\sim t_m^{-1}.
\end{equation} 
We thus conclude that $\eta=1$, for all processes with correlation function in the form of Eq.~(\ref{eq:ScaledCorr}).
%Note that experimentalists report $1/f$ noise up to the lowest measurable frequency  $t_m^{-1}$ (according to traditional definitions) hence increasing measurement time does not yield new insights if the underlying lying process is truly scale invariant, and in this sense looking for the low-frequency cutoff of $1/f$ can be futile.
 
The existence of such a cutoff is required since a purely $1/f$ noise cannot exist in the range $0<f<\infty$ for the following reasons; first, the power at zero frequency must be finite since $\left\langle \bar{I}_{t_m}^2\right\rangle <\infty$, for every finite measurement time. Second,  we expect the total power to be finite at every finite measurement time, since
\begin{equation}
\int_{-\infty}^{\infty}{\rm d}\omega\left\langle S_{t_m}(\omega)\right\rangle=2\pi t_m^{\Upsilon}\varphi_{\rm TA}(0),
\label{eq:Total}
\end{equation}
even though Eq.~\eqref{eq:NonStationatyPS} is not integrable in $[0,\infty)$. Notice that the sample spectrum is time-dependent although its $1/f$ part might be time-independent, e.g. if $\Upsilon={\rm V}$. Therefore, measuring $1/f$ noise, even time independent, does not contradict the finite power requirement. Equation \eqref{eq:Total} implies that $\langle S_{t_m}(\omega)\rangle$ is normalized, provided $\varphi_{\rm TA}(0)$ is finite, and since it is non-negative it satisfies the conditions for a normalized density. 

%\begin{widetext}

\begin{table*}	
\center{
	\begin{tabular}{|c|c|c|c|c|}
	\hline &  & $\Upsilon$  & ${\rm V}$ & $\Lambda$  \\ \hline
	 single file diffusion & & $1/2$ & $1/2$ & -1/2   \\ \hline
	 blinking quantum dot - finite mean `on' time& $ 0<\alpha<1$  & $2\alpha-2$ & $\alpha-1$ &  $\alpha-1$ \\ \hline
	 blinking quantum dot - infinite mean & $ 0<\alpha<1$ & $0$ & $1-\alpha$ &  $-\alpha$ \\ \hline
	 logarithmic potential & $ 1<\alpha<3$ & $2-\alpha$ & $2-\alpha$ & $1/2-\alpha$ \\ \hline	 
		\end{tabular}
		\caption{Summary of the scaling correlation function exponents (see Eq.~\eqref{eq:ScaledCorr1}) for the three systems discussed in sections VI-VIII. The first and second columns list the system and its relevant scaling exponent range.} 
		\label{Tab2}}
\end{table*}
%\end{widetext}

%\begin{widetext}

\begin{table*}	\center{
	\begin{tabular}{|c|c|c|c|c|c|c|}
	\hline &  & $\beta$  & $z$ & $\eta$ & $\mu$ & $\delta$  \\ 
				 &  & $S\sim \omega^{-\beta}$ & $S\sim t_m^{-z}$ & $\omega_{min}\sim t_m^{-\eta}$ & $S(0)\sim t_m^{\mu}$ & $\int_{1/t_m}^{\omega_{max}}S_{t_m}(\omega){\rm d}\omega$ \\\hline
	 single file diffusion & & $3/2$ & $0$ & $1$ & $3/2$ & $1/2$ \\ \hline
	 blinking quantum dot - finite mean `on' time& $ 0<\alpha<1$  & $\alpha$ & $1-\alpha$ & $1$ & $2\alpha-1$ & $\alpha-1$ \\ \hline
	 blinking quantum dot - infinite mean & $ 0<\alpha<1$ & $2-\alpha$ & $1-\alpha$ & $1$ & $1$ & $0$\\ \hline
	\multirow{2}{*}{logarithmic potential} & $ 1<\alpha<2$ & \multirow{2}{*}{$3-\alpha$} & \multirow{2}{*}{$0$} & \multirow{2}{*}{$1$} & \multirow{2}{*}{$3-\alpha$} & $2-\alpha $ \\ \cline{2-2} \cline{7-7}
	                       & $ 2<\alpha<3$ &  &  &  & & 0 \\ \hline	 
		\end{tabular}
		\caption{Summary of the critical exponents for the three systems discussed in sections VI-VIII. }
		\label{Tab3}}
\end{table*}
%\end{widetext}

\subsection{The Total Power Time Dependence}
For an ideal $1/f$ source the total power of the process diverges since 
$\int_0 ^\infty \omega^{-\beta} d \omega=\infty$. If $\beta <1$ ($\beta>1$) the integral diverges due to the high (low) frequency limit. We should be mainly concerned with low frequency, since the whole behavior of $1/f$ is found in that regime. Indeed, as we show below,  there always exists a physical mechanism that leads to cutoff at large frequencies. 
As was mentioned in Sect. \ref{CriticalExponents} the total measured power is characterized by the exponent $\delta$, i.e. $\int_{1/t_m}^{\omega_\text{max}}S(\omega){\rm d}\omega\sim t_m^{\delta}$, where we assume that $\omega_\text{max}$ is time independent. 
As a result
\begin{equation}
\int_{1/t_m}^{\omega_\text{max}}S(\omega){\rm d}\omega\sim\omega_{\rm max}^{-{\rm V}}t_m^{\Upsilon-{\rm V}}+t_m^\Upsilon=\omega_{\rm max}^{-{\rm V}}t_m^{-z}+t_m^{\mu-1}.
\end{equation}
The exponent $\delta$ is accordingly given by 
\begin{equation}
\delta=-\min(z,1-\mu).
\end{equation}
%Taking the lower integration boundary to zero does not affect the time dependence of the spectrum, however by 
Taking $\omega_{\text max}\rightarrow\infty$ the total power time dependence scales as $t_m^{\Upsilon}$ as is expected in Eq.~\eqref{eq:Total}. 

For bounded processes, we expect that the total power will not increase as a function of time, namely $\Upsilon\leq 0$. By contrast, in open (i.e. unbounded) processes, the total measured power may increase with measurement time, as for example single-file diffusion in infinite system, as we will discuss below.

\section{The basic requirements for the power spectrum}
\label{Sec6}
We now compare between the properties of Wienerian and aging power spectra
clarifying the meaning of the latter.
Stationary processes $I(t)$ and their power spectrum have the following properties:
\begin{itemize}
\item[(i)] $\langle I \rangle$ is a constant independent of time.
\item[(ii)] $\langle I^2 \rangle$ is a constant independent of time and the correlations function $\langle I(t+\tau) I(t) \rangle$ is a function of $\tau$ only. 
\item[(iii)] The power spectrum is non negative.
\item[(iv)] The total power is 
\begin{equation}
P_T = \int_{-\infty} ^\infty S(\omega) {\rm d} \omega = 2 \pi \langle I^2 \rangle.
\end{equation}
This well-known property is easily verified using the Wiener-Khinchin theorem.
\item[(v)] The total power is
\begin{equation}
P_T = 2 \pi  \sum_{n=-\infty} ^\infty \langle |a_n|^2 \rangle =
 2 \pi \langle I^2 \rangle
\label{eqPITA}
\end{equation} 
where $a_n$ is the Fourier amplitude $a_n= \int_0 ^{t_m} \exp( - i \omega_n t) I(t) {\rm d} t/t_m$
and $\omega_n = 2 \pi n /t_m$. Here $\langle S(\omega)\rangle= t_m |a_n|^2$ in the limit
of large $t_m$. 
\end{itemize}
See further discussion in Appendix \ref{sec:Fourier} and in Ref.~\cite{Kubo}.
Properties $(iii-v)$ are important, they show that the power spectrum is
the distribution of the modes of the system, and that we may normalize
the power spectrum. Indeed in many cases the normalized power spectral density
is considered, namely $\langle S(\omega) \rangle/ [ 2 \pi \langle I^2 \rangle]$.

Our approach provides a non-Wienerian framework for the power spectrum.
It is thus natural to ask how the above points translate to the nonstationary, non-Wienerian case.
\begin{itemize}
\item[(i)] $\langle I \rangle$ is a constant independent of time.
\item[(ii)] Here $\langle I(t+\tau) I (t) \rangle = t^\Upsilon \phi_{{\rm EA}} (\tau/t)$
is the starting point. So we get by definition 
\begin{equation}
\langle I^2(t)\rangle = t^\Upsilon \phi_{{\rm EA}}(0).
\label{eq123}
\end{equation}
\item[(iii)]
The power spectrum $\langle S_{t_m} (\omega) \rangle \ge 0$
due the definition of the periodogram. 
\item[(iv)] We find
\begin{multline}
\int_{-\infty} ^\infty  \langle S_{t_m}  \left(\omega\right) \rangle{\rm d} \omega =
\frac{2 \pi}{t_m} \left\langle  \int_0 ^{t_m}  I^2 (t) {\rm d} t\right\rangle=\\
=2 \pi \left\langle  \overline{I^2} \right\rangle= 2 \pi \varphi_{{\rm TA}}(0).
\label{eq321}
\end{multline}
Here the overline stands for a time average. 
The key to the interpretation of the power-spectrum is Eq. (13)
from which we find
\begin{equation}
\varphi_{{\rm TA}}(0) =\lim_{x \to 0} x^{1 + \Upsilon} \int_x ^\infty {\rm d} y \frac{\phi_{{\rm EA}} (y)}{y^{2 + \Upsilon} } .
\end{equation}
Using L'H\^opital's rule we find
\begin{equation}
\varphi_{{\rm TA}} (0) = \frac{\phi_{{\rm EA}} (0)}{1 +\Upsilon}.
\end{equation}
Thus using Eqs.  
(\ref{eq123},\ref{eq321}) the total power is
\begin{equation} 
P_T = \int_{-\infty} ^\infty S_{t_m} (\omega) {\rm d} \omega = 
\frac{ 2 \pi \langle I^2 \rangle}{1 + \Upsilon} . 
\end{equation} 
Hence exactly like the stationary theory, for scale invariant correlation functions,
the total power is given by $\langle I^2 \rangle$.
 Hence it is fully justified to call $\langle S_{t_m}(\omega)\rangle$ the power spectral density. 
When $\Upsilon=0$ the analogy
is complete. 
\item[(v)] Also here rule   
(\ref{eqPITA}) holds, with $P_T$ given in the  previous item (see discussion
in Appendix \ref{sec:Fourier} with respect to filtering).
\end{itemize}
We see that even though the correlation function is by definition 
far from being stationary, the main structure of power spectrum theory is left untouched,
though now the power spectrum is dependent on the measurement time. 

Since the aging spectrum is very different from the Wienerian one, we actually have two methods to present it (as shown all along this work).
The spectrum of $1/f$ noise source of the type discussed in this manuscript, should be presented, if possible, using two plots; The first is $\langle S(\omega)\rangle$ versus $\omega$ where $\omega_n= 2\pi n/t_m$, this represents the {\em true spectrum} in the sense that Fourier modes in $(0,t_m)$ are discrete. As well known within this traditional presentation, $1/f$ noise presents an infrared divergence, $S(f)\sim f^{-\beta}$ which $\beta\geq 1$ implies naively that the total energy of the process is infinite (the low-frequency paradox of $1/f$ noise \cite{Mandelbrot1967,Niemann}). To gain better insight on this low-frequency behavior we consider a second ``spectrum'' where $\langle S(\omega)\rangle$ is continuous in $0\leq\omega<\infty$. This spectrum %is evaluated from computer data, and it 
yields insights on the low-frequency behavior, namely the oscillations of the spectrum and the zero-frequency component $\langle S(0)\rangle$. All these yield significant insight on the process, e.g. the exponents $\Upsilon$, ${\rm V}$, $\Lambda$, $\beta$, $z$, $\eta$, $\mu$ and $\delta$. By contrast, evaluating the spectrum at the natural frequencies yields only the $1/f$ component, which provides partial information on the spectrum, i.e. the exponents $\Upsilon$, $\beta$, $z$ and $\delta$.  We would like to emphasize that all
five exponents related to the spectrum; $\beta$, $z$, $\eta$, $\mu$ and $\delta$ can be evaluated
from natural frequency data, provided that one adds one single measurement:  the time dependent 
behavior of $S_{t_m}(0)$. And three of them: $\beta$, $z$, and $\delta$ can be evaluated even without information on the zero frequency spectrum (then the cutoff is simply $1/t_{m}$).  The continuous spectrum advantage is that it can give 
further information, for example by the analysis of oscillations of the spectrum we may estimate  $\Lambda$ which yields information on the large
argument behavior of the  correlation function. So while $\langle S(\omega)\rangle$ on the natural frequencies $\omega_n=2\pi n/t_m$ has the advantage of a clear interpretation in terms of Fourier modes, see e.g. Kubo et al. \cite{Kubo}, the continuous spectrum gives insights on the underlying process which should not be ignored. 
%Apart from generalizing the \WK to a large class of aging systems, our work thus gives meaning to the spectrum as a continuous function of $\omega$. This reveals new features in the spectrum, which are absent when sampling at discrete frequencies $\omega_n = 2 \pi n/t_m$, and which can be related to the behavior of the autocorrelation function. We further showed how the time-dependence of the spectrum leads to a finite power for all measurement times. 
The fact that the scaling correlation function on which we base our analysis can be observed in a wide range of systems, beyond the specific examples discussed here, underlines the universality of our main results.

\section{Blinking Quantum Dot Model}
\label{BlinkingQD}
We next demonstrate the above results by studying a stochastic model for blinking-quantum dots. A quantum dot is a nanocrystal that, when interacting with a continuous wave exciting laser field, switches at random times between on and off states \cite{Kuno,PhysTodayBarkai}. Such a process is an example of a renewal process. To analyze such systems we follow previous works (e.g. \cite{Margolin04,Margolin06,Margolin05}) and define a two-state system, where $I(t)=0$ is the state ``off'' and $I(t)=1$ is ``on''. Without loss of generality we choose the system to be initially in $I(0)=1$. At each time $t_n$ the system switches to the other state alternately (``on'' $\rightarrow$ ``off'' or ``off'' $\rightarrow$ ``on''). The renewal times are $t_n=\sum_0^n\tau_i$ where $\left\{\tau_i\right\}$ are distributed according to the PDF $\psi(\tau)$ and $n$ is the number of renewals until time $t_n$ (see Fig. \ref{fig:BQD.Model}).
We assume that the ``on'' and ``off'' times $\left\{\tau_i\right\}$ are uncorrelated. The ``off'' sojourn times are power-law distributed $\psi(\tau)\sim \tau^{-(1+\alpha)}$ with $0<\alpha<1$. For the ``on'' times we consider two cases; In the first one the ``on'' times are distributed with infinite mean, e.g. power-law distribution $\psi(\tau)\sim \tau^{-(1+\alpha)}$ (see realization of $I(t)$ in Fig. \ref{fig:trajectories} with $\alpha=0.8$). In the second case we consider ``on'' times with finite mean distribution, e.g. power-law distribution with an exponentially decaying of the tail. Both cases were experimentally examined \cite{Sadegh,Pelton}. 
Such a system follows a power-law intermittency route to $1/f$ noise. This means that power-law waiting times in a substate of the system are responsible for the observed spectrum. This approach was suggested as a
fundamental mechanism for $1/f$ noise in the context of intermittency of chaos and turbulence in the work of
Manneville \cite{Manneville}. We note that the renewal process describes not only blinking dots, but also the trap model, a well-known model of glassy dynamics \cite{Bouchaud96,Dean}. The connection between the two systems is the power-law waiting times in the microstates of the system.

\begin{figure}
	\centering
		\includegraphics[width=0.9\columnwidth]{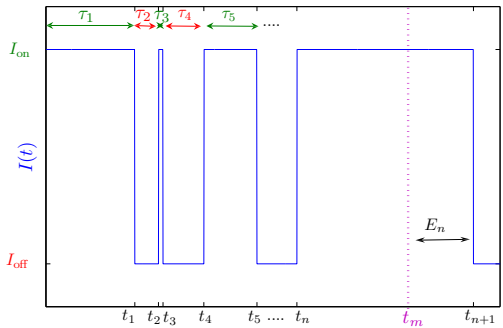}
		\caption{A single realization of the signal $I(t)$ versus time (blue). $\left\{\tau_i\right\}$ are the sojourn times at each state; ``on'' (green) and ``off'' (red). $\left\{t_i\right\}$ are the renewal times. $t_m$ is the measurement time and $E_n$ refers to the forward recurrence time.}
	\label{fig:BQD.Model}
\end{figure}

\subsection{Infinite Mean ``On'' Sojourn Time Distribution}
\subsubsection{Low-Frequency Power Spectrum}
In the model both ``on'' and ``off'' times are power-law distributed
\begin{equation}
\psi_{\text{off/on}}(\tau)\sim (\tau_0/\tau)^{1+\alpha}
\label{eq:LongTailDistribution}
\end{equation}
where $\tau_0$ is a microscopic time scale, $\tau>\tau_0$ and $0<\alpha<1$ (e.g. see the experiment in \cite{Pelton}). We choose for both substates, ``on'' and ``off'', the same exponent $\alpha$ for simplicity. Typical values of $\alpha$ in experiments are $0.5< \alpha<0.8$.
This case was studied analytically before in Refs. \cite{Lowen,Margolin06,Margolin05,Margolin04,Niemann,Godreche}.    

%In a nonstationary process, the correlation function $C(t,\tau)$ and the time-averaged correlation function $\left\langle C_{TA}(t,\tau)\right\rangle$ are not identical as mentioned. %To demonstrate this issue we plot $C(t,\tau)$ and $\left\langle C_{TA}(t,\tau)\right\rangle$ for the case where ``on'' and ``off'' times are distributed with long tail (Eq. (\ref{eq:LongTailDistribution})) and $\alpha=0.5$ (see Fig. \ref{fig:Corr_TACorr}). 
The analytic formulas for the time- and ensemble-average correlation function are given in \cite{Margolin06}, where $t$ and $\tau$ are larger then $\tau_0$, positive and comparable;
\begin{eqnarray}
&&\phi_{\rm EA}(x)=\frac{1}{2}-\frac{1}{4}\frac{\sin(\pi\alpha)}{\pi}B\left(\frac{x}{1+x};1-\alpha,\alpha\right) \label{eq:CorrBQD}\\
&&\varphi_{\rm TA}(x)=  \nonumber \\ &&\frac{1}{4}+\frac{1}{4}\frac{\sin(\pi\alpha)}{\pi}\left[\frac{B\left(1-x;\alpha,1-\alpha\right)}{1-x}-\frac{1}{\alpha}\left(\frac{x}{1-x}\right)^{1-\alpha}\right],  \nonumber
\end{eqnarray}
where $x=\tau/t$ and $B(z;a,b)\equiv \int_0^z{\rm d}x(1-x)^{b-1}x^{a-1}$ is the incomplete Beta function. In order to determine the power spectrum for this nonstationary process we need to use the aging \WK Eqs. (\ref{eq:GenerlizedWK}) or (\ref{eq:InfiniteCovarianceDensity}) instead of Eq.~(\ref{eq:WK}).

%\begin{figure}
%	\centering
%		\includegraphics[width=0.95\columnwidth]{C:/MATLAB701/work/Corr_TACorr.eps}
%	\caption{The correlation function $C(t,\tau)$ (green curve) and the time averaged correlation function $\left\langle C_{TA}(t,\tau)\right\rangle$ (blue curve) versus the ratio $\tau/t$ is a case where both ``on'' and ``off'' sojourn times are power-law distributed with exponent $\alpha=0.5$.  }
%		\label{fig:Corr_TACorr}
%\end{figure} 

\begin{figure}
	\centering
	\includegraphics[width=0.95\columnwidth]{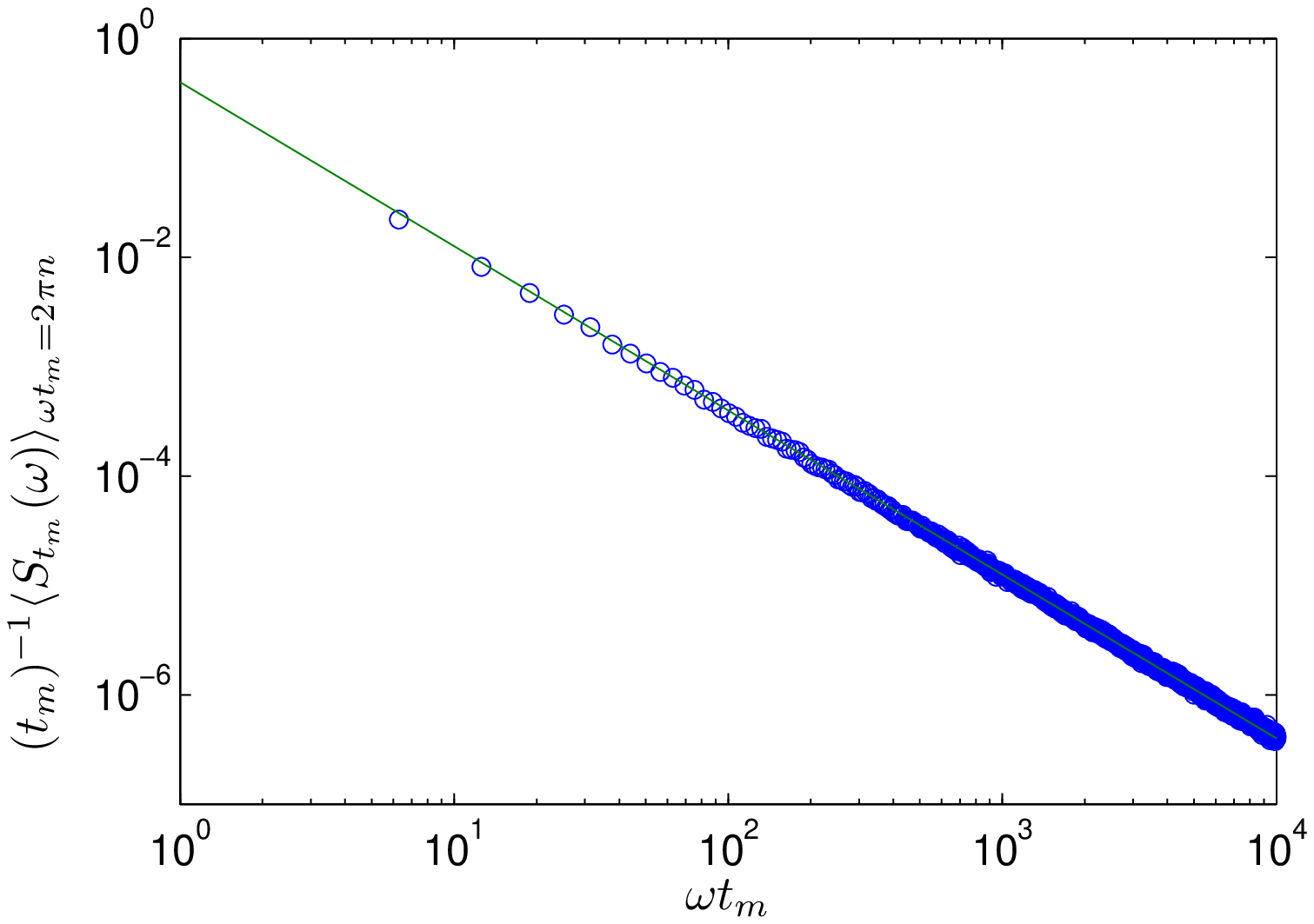}
		\includegraphics[width=0.95\columnwidth]{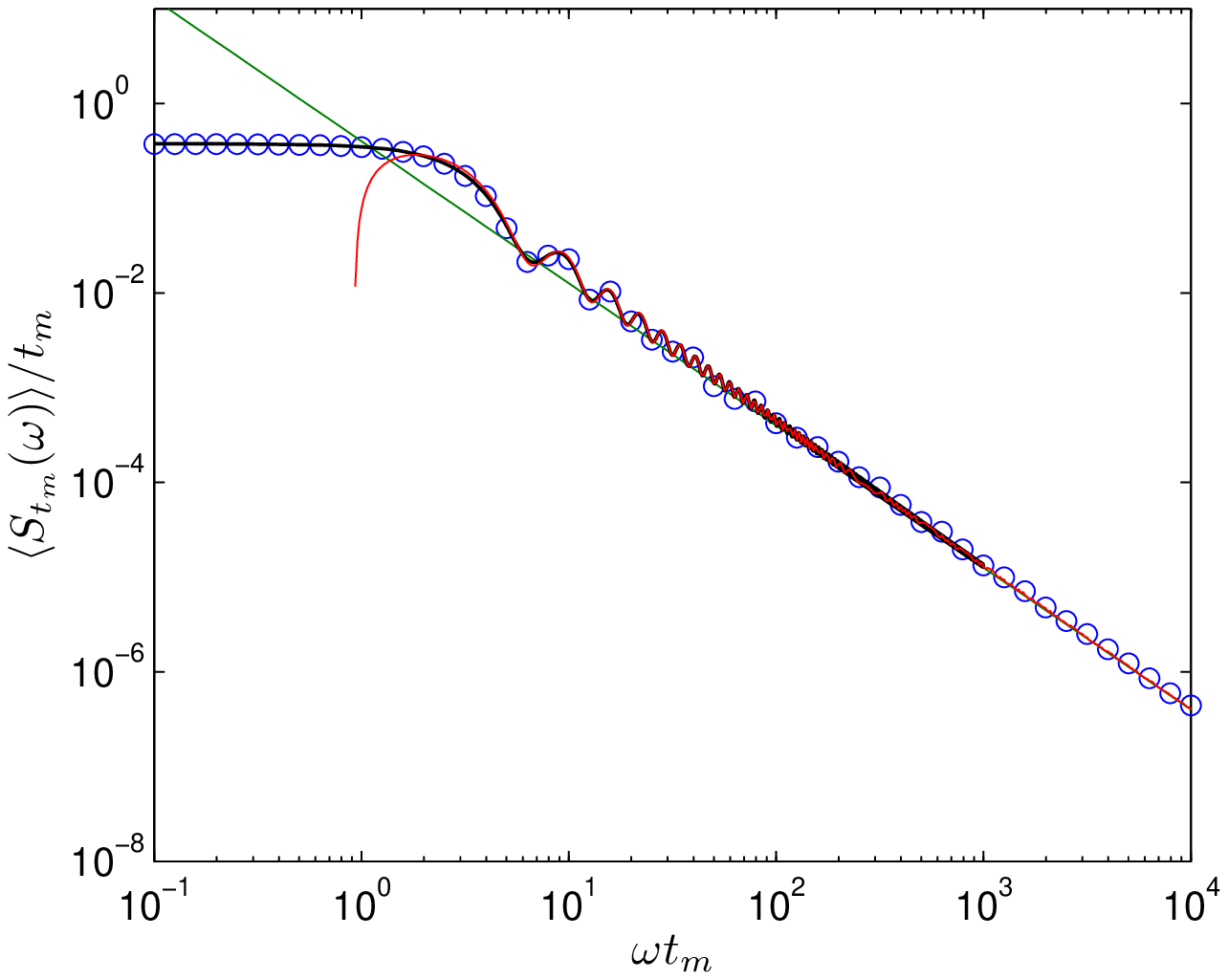}		
		\caption{Simulation results for a blinking quantum dot model where both distributions of ``on'' and ``off'' sojourn times are fat tailed Eq.~\eqref{eq:LongTailDistribution} when $\alpha=0.5$ at measurement time $t_m=10^5$. In the upper panel the spectrum is given in natural (discrete) frequencies $\omega t_m=2\pi n $ and compare to Eq.~(\ref{eq:PS_inf}) represented in green line. In the lower panel the power spectrum is taken in continuous frequencies where the red line represents limit behavior \eqref{eq:PS_inf_osc} and the entire analytic prediction is presented in black line Eq.~(\ref{eq:ICD.BQD}). The simulations method is given in App. C.}
	\label{fig:PS_alpha0.5}
\end{figure}

%\begin{figure}
%\centering
%	\includegraphics[width=0.95\columnwidth]{../../../../MATLAB701/work/PS.alpha.0.5.ICD.eps} \\
%%	\includegraphics[width=0.8\columnwidth]{C:/MATLAB701/work/PS.OnOff.alpha05.eps}
%	\caption{Simulation results in $\log_{10}$-$\log_{10}$ scale for the $\left\langle S_{t_m}(\omega)\right\rangle$ when $\alpha=0.5$ for three different measurement times, $t_m=10^2$ (blue), $t_m=10^4$ (red) and $t_m=10^6$ (green). 
% The black solid line is Eq. (\ref{eq:NonNormalized}). Notice that x-axis is $\omega$, while in Fig. \ref{fig:PS_alpha0.5} it is $\omega t_m$.}
%		\label{fig:PS.alpha.0.5.ICD.eps}
%\end{figure}

We find by using Eqs.~(\ref{eq:InfiniteCovarianceDensity}) and (\ref{eq:CorrBQD}) (see derivation in App. \ref{Exact})
\begin{equation}
\langle S_{t_m}(\omega)\rangle/t_m=\frac{1}{4}{\rm sinc}^2\left(\frac{\tilde{\omega}}{2}\right)+ \frac{1}{2\tilde{\omega}}\Im\left[M(1-\alpha,2;\imath\tilde{\omega})\right],
\label{eq:ICD.BQD} 
\end{equation}
where $\tilde{\omega}=\omega t_m$, $M(a,b;x)$ is the Kummer confluent hypergeometric
function and $\Im[.]$ refers to its imaginary part. The ${\rm sinc}^2(.)$ term is the contribution to the spectrum from a constant. Equation \eqref{eq:ICD.BQD} predicts the behavior of the power spectrum where $t_m\rightarrow\infty$ but $\tilde{\omega}$ remains finite.

%\subsubsection{High-Frequency Power Spectrum}
%We note that Eq. (\ref{eq:ICD.BQD}) is valid only for finite $\tilde{\omega}$ even though Eq. (\ref{eq:InfiniteCovarianceDensity}) imposes no such restriction. As illustrated in Fig. \ref{fig:PS.alpha.0.5.ICD.eps}, a different behavior emerges at large $\omega$. This is a consequence of taking the correlation function in long time limit, i.e. $t_m$,$\tau$ $\rightarrow\infty$. Information about the correlation function for short $\tau$ is necessary to find the behavior of the spectrum at high frequencies.
%
%For $\omega t_m \gg 1$ and arbitrary $\omega$, we find
%\begin{eqnarray}
%&&\langle S_{t_m}(\omega)\rangle_{\omega>}=  \frac{t_m^{\alpha-1}}{2a\omega^2\Upsilon(\alpha+1)}\times  \label{eq:NonNormalized} \\
%&&\left[1-\psi_{on}(\imath\omega)-\frac{\left[1-\psi_{on}(\imath\omega)\right]^2\psi_{off}(\imath\omega)}{1-\psi_{on}(\imath\omega)\psi_{off}(\imath\omega)}+c.c.\right],   \nonumber
%\end{eqnarray}
%where we note that the spectrum depends on the details of the process, namely it depends on $\psi(\imath\omega)$. 
%The above spectral density is non-normalized, since in $\omega \rightarrow 0$ it behaves as $\omega^{-2+\alpha}$. A more detailed discussion on such spectral densities will be published elsewhere \cite{Notation}. Simulations are compared to the analytic prediction Eq. (\ref{eq:NonNormalized}) in Fig. \ref{fig:PS.alpha.0.5.ICD.eps}; there we use $\psi(\imath\omega)=a  E(1+\alpha,\imath\omega)$ where $E(n,z)=\int_1^{\infty}\exp(-zt)t^{-n}{\rm d}t$ is the exponent integral.
%

\subsubsection{$1/f$ Noise}
The average power spectrum of the signal $I(t)$, 
by using Eqs.~(\ref{eq:ScaledCorr1},\ref{eq:1FNoise},\ref{eq:CorrBQD}), is
\begin{equation}
\langle S_{t_m}(\omega) \rangle_{\omega t_m=2\pi n} \approx \frac{\cos(\alpha\pi/2)}{2\Gamma(1+\alpha)}t_m^{\alpha-1}\omega^{\alpha-2}.
\label{eq:PS_inf}
\end{equation}	 
The same result is found by taking the limit of $\omega t_m \gg 1$ in Eq.~(\ref{eq:ICD.BQD}). % or equivalently taking the limit of small $\omega$ in \eqref{eq:NonNormalized}, i.e. in the $1/f$ regime the two solutions, Eqs. (\ref{eq:ICD.BQD},\ref{eq:NonNormalized}) match. 
The aging \WK reproduces the result Eq.~\eqref{eq:PS_inf} that was found before, e.g. in Ref.~\cite{Margolin06}. 
To evaluate the oscillating behavior we use Eq.~\eqref{eq:NonStationatyPS} and find
\begin{eqnarray}
&&\left\langle S_{t_m}(\omega)\right\rangle_{\omega t_m \gg 1}\approx \frac{\cos(\alpha\pi/2)}{2\Gamma(1+\alpha)}t_m^{\alpha-1}\omega^{\alpha-2} \label{eq:PS_inf_osc}  \\
&&+\frac{t_m}{4}{\rm sinc}^2\left(\frac{\omega t_m}{2}\right)-\frac{\cos\left[\omega t_m-\alpha\pi/2\right]}{2\Gamma(1-\alpha)\omega^{2+\alpha}t_m^{\alpha+1}}.
\nonumber
\end{eqnarray}
In Fig. \ref{fig:PS_alpha0.5} simulation results are compared with the exact analytic prediction Eq.~\eqref{eq:ICD.BQD} and excellent agreement is observed. Figure \ref{fig:PS_alpha0.5} further confirms the validity of the two approximated spectra: First, the $1/f$ noise Eq.~\eqref{eq:PS_inf} for discrete frequencies $\omega t_m=2\pi n$ agrees well with the simulations when $n\in {\mathbb N}$. Moreover, the oscillatory behavior  Eq.~\eqref{eq:PS_inf_osc} for continuous frequencies, presents a good agreement in the limit $\omega t_m \gg 1$. %Those oscillations are a finite-measurement-time-effect and their amplitude is determined by the behavior of the correlation function where $\tau \sim t_m$ as is predicted in Eq. \eqref{eq:PS_inf_osc}

\subsubsection{Critical Exponents}

The critical exponent $\beta$, which is determined by the power-law decay of the average power spectra, is  $\beta=2-\alpha$ for the infinite mean `on' sojourn time. The aging exponent which is related to the time decay is  $z=1-\alpha$.

The averaged zero-frequency power spectrum is defined as $\left\langle S(\omega=0)\right\rangle=t_m^{-1}\int_0^{t_m}{\rm d}t_1\int_0^{t_m}{\rm d}t_2\left\langle I(t_1)I(t_2)\right\rangle$.
Using Eq.~\eqref{eq:CorrBQD} we find 
\begin{equation}
\left\langle S_{t_m}(0)\right\rangle=\frac{1}{4}(2-\alpha)t_m 
\label{eq:St0}
\end{equation}
hence, its related exponent is $\mu=1$.
The transition point between $\left\langle S_{t_m}(\omega)\right\rangle$ when $\omega\neq 0$ to the behavior at $\left\langle S_{t_m}(\omega=0)\right\rangle$ is defined as the low-frequencies cutoff $\omega_{\rm min}$;
\begin{equation}
\omega_{\rm min}=\left(\frac{2\cos(\alpha\pi/2)}{\Gamma(1+\alpha)(1-\alpha)}\right)^{\frac{1}{2-\alpha}}t_m^{-1}. \label{eq:OmegaMin}
\end{equation}
Therefore, the exponent $\eta=1$.

The highest frequency where $1/f$ appears, $\omega_{\rm max}$, is related to frequency at which the approximation \eqref{eq:ScaledCorr1} fails. Using the  Laplace transform of the waiting time PDF $\hat{\psi}(u) \sim 1 - \Gamma(1-\alpha)(\tau_0 u)^\alpha$, such approximation holds when $\omega\ll\omega_{\rm max}\sim \tau_0^{-1}$.
The behavior of the total measured power is eventually
\begin{equation}
\int_{\omega_{\rm min}}^{\omega_{\rm max}}{\rm d}\omega\left\langle S_{t_m}(\omega)\right\rangle \propto Const.
\end{equation}
We hence find $\delta=0=-{\rm min}(z=1-\alpha,1-\mu=0)$ as expected.

\subsection{Finite Mean ``On'' Sojourn Time Distribution}
%\subsubsection{The Power Spectrum for Low and High Frequencies}
Now we consider that the ``on'' sojourn times have a finite mean $\left\langle \tau\right\rangle$, while the ``off'' times are power-law distributed as the first case $\psi_{\rm off}(\tau)\sim (\tau_0/\tau)^{1+\alpha}$ (see the measurements in \cite{Sadegh}). In Laplace space ($s\rightarrow t$) we find for small $s$ when $0<\alpha<1$
\begin{eqnarray}
\hat{\psi}_{\rm on}(s)&=&1-\langle\tau\rangle s + ... \\ \nonumber
\hat{\psi}_{\rm off}(s)&=&1-a s^{\alpha} + ...
\end{eqnarray}
where $a=\Gamma(1-\alpha)\tau_0^{\alpha}$.
The ensemble average correlation function in the limit $\tau,t\rightarrow\infty$ scales as \cite{Margolin04}
\begin{equation}
C(t,\tau)=\frac{\left\langle \tau\right\rangle^2}{a^2\Gamma^2(\alpha)}(t\cdot\tau)^{\alpha-1}.
\label{eq:CorrFin}
\end{equation}
Using Eqs.~(\ref{eq:C3}) and (\ref{eq:GenerlizedWK}), we obtain the power spectrum
\begin{eqnarray} 
\label{eq:PS_fin}
\left\langle S_{t_m}(\omega)\right\rangle=&&\frac{2\left\langle \tau\right\rangle^2\pi\Gamma(1-\alpha)}{\alpha a^24^{\alpha}\Gamma(\alpha)}\cdot   \\
&&{_2\tilde{F}_3}\left[\frac{1+\alpha}{2},\frac{\alpha}{2};\frac{1}{2},\frac{1}{2}+\alpha,1+\alpha; -\frac{\tilde{\omega}^2}{4}\right],\nonumber
\end{eqnarray}
where ${_2\tilde{F}_3}\left[a_1,a_2;b_1,b_2,b_3;z\right]$ is the regularized hypergeometric function (see Fig. \ref{fig:PS.OnOff.alpha0.5}).
%Eq. \eqref{eq:PS_fin} predicts the measured spectrum only at low frequencies ($\omega \ll 1$).
%For high frequencies we find (see App. \ref{Exact})
%\begin{eqnarray}
%&&\langle S_{t_m}(\omega)\rangle_{\omega>}=  \frac{t_m^{\alpha-1}}{a\omega^2\Upsilon(\alpha+1)}\cdot  \label{eq:NND_Fin}\\
%&&\cdot\left[1-\psi_{on}(\imath\omega)-\frac{\left[1-\psi_{on}(\imath\omega)\right]^2\psi_{off}(\imath\omega)}{1-\psi_{on}(\imath\omega)\psi_{off}(\imath\omega)}+c.c.\right],   \nonumber
%\end{eqnarray}  
%%
%where now we use $\psi_{on}(\imath\omega)\sim 1-\imath\omega\langle\tau\rangle$.

We note that Eqs. (\ref{eq:ICD.BQD}) and \eqref{eq:PS_fin} are valid only for finite $\tilde{\omega}$ even though Eq.~(\ref{eq:InfiniteCovarianceDensity}) imposes no such restriction. For very large frequency we expect to get non-scaling deviations, since then the scale invariant correlation function is not strictly valid for the Wiener-Khinchim theorem. For illustrations see Fig.~\ref{fig:PS.OnOff.alpha0.5} where a different behavior emerges at large $\omega$. This is a consequence of taking the correlation function in long time limit, i.e. $t$,$\tau$ $\rightarrow\infty$. Information about the correlation function for short $\tau$ is necessary to find the behavior of the spectrum at high frequencies. A more detailed discussion will be published elsewhere \cite{Notation}. However, Fig.~\ref{fig:PS.OnOff.alpha0.5} clearly illustrates that as we increase the measurement time the asymptotic theory perfectly matches the theory.

\subsubsection{$1/f$ Noise}
By using Eqs.~(\ref{eq:ScaledCorr1}},\ref{eq:1FNoise},\ref{eq:CorrFin}) we obtain 
\begin{equation}
\langle S_{t_m}(\omega)\rangle_{\omega t_m =2\pi n} \approx \frac{2\langle\tau\rangle^2\cos(\pi\alpha/2)}{\Gamma(\alpha)a^2\alpha}t_m^{\alpha-1}\omega^{-\alpha}.
\label{eq:PS_fin_1F}
\end{equation}
As in the infinite mean ``on'' times, the same result is found by taking the limit $\tilde{\omega}\gg 1$ in Eq.~(\ref{eq:PS_fin}). %, or $\omega \rightarrow 0$ in Eq. (\ref{eq:NND_Fin}), i.e. the two solutions match in the intermediate regime.
In addition, we examine the oscillating behavior
\begin{eqnarray}
\langle S_{t_m}(\omega)\rangle_{\omega t_m \gg 1} &&\approx \frac{2\langle\tau\rangle^2\cos(\pi\alpha/2)}{\Gamma(\alpha)a^2\alpha}t_m^{\alpha-1}\omega^{-\alpha}\label{eq:PS_fin_osc}  \\
&&+\frac{2\langle\tau\rangle^2\Gamma(1+\alpha)}{(1+\alpha)\Gamma^2(\alpha)a^2}\frac{\sin\left(\omega t_m-\alpha\pi/2\right)}{\omega^{1+\alpha}t_m^{\alpha}}.
\nonumber
\end{eqnarray}
In Fig. \ref{fig:PS.OnOff.alpha0.5} we compare the simulation results with the theory Eq. \eqref{eq:PS_fin}. The $1/f$ noise with oscillatory corrections Eq. \eqref{eq:PS_fin_osc} shows good agreement with numerical results for intermediate frequencies.

\subsubsection{Critical Exponents}
From Eq.~\eqref{eq:PS_inf} we conclude that $\beta=\alpha$ and $z=1-\alpha$.
The averaged zero-frequency power spectrum is then
\begin{equation}
\left\langle S_{t_m}(0)\right\rangle=\frac{\left\langle \tau\right\rangle^2}{\alpha\Gamma(2\alpha)a^2}t_m^{2\alpha-1}.
\label{eq:St0offon}
\end{equation}
We therefore find $\mu=2\alpha-1$ for this case. This result is reasonable since the zero frequency power is the squared-time-average signal $I(t)$ times the measurement time. The integral over $I(t)$ between zero and $t_m$ is proportional to $N_{t_m}$ times the average ``on'' time $\left\langle \tau\right\rangle$, where $N_{t_m}$ is the number of renewals until time $t_m$. $N_{t_m}$ itself is proportional to $t_m^{\alpha}$, a well known result in renewal theory \cite{Godreche}.
Hence we get $S_{t_m}(0)\propto t_m^{2 \alpha -1}$.

Interestingly in this case the value of $\alpha$ changes the behavior of $\left\langle S_{t_m}(0)\right\rangle $, namely if $\alpha>1/2$ then $\left\langle S_{t_m}(0)\right\rangle$ increases in time, if $\alpha<1/2$, it decreases in time, and when $\alpha=1/2$ then $\left\langle S_{t_m}(0)\right\rangle$ is time independent.

The low-frequencies cutoff $\omega_{\rm min}$ is,
\begin{equation}
\omega_{\rm min}=\left(\frac{2\cos(\pi\alpha/2)\Gamma(2\alpha)}{\Gamma(\alpha)}\right)^{1/\alpha}t_m^{-1}. 
\label{eq:OmegaMinOffOn}
\end{equation}
Therefore we find that in both cases the low-frequencies cutoff decays in time with the same exponent, $\eta=1$, and differs by a prefactor only.

The behavior of the total measured power is additionally
\begin{equation}
\int_{\omega_{\rm min}}^{\omega_{\rm max}}{\rm d}\omega\left\langle S_{t_m}(\omega)\right\rangle \propto t_m^{\alpha-1}.
\end{equation}
As a result $\delta=\alpha-1$. The decrease of the total power with measurement time is reasonable since the signal exhibits longer and longer ``off'' times, while the ``on'' times remains finite. 

In Eq.~\eqref{eq:Total} we showed that the total power proportional to $t_m^{\gamma}\varphi_{TA}(0)$. This is true for the ideal process, where the scale invariant correlation functions are valid for all $t$ and $\tau$, and then $\varphi_{TA}(0)<\infty$. We note that Eq.~\eqref{eq:CorrFin} gives $\varphi_{TA}(0)\rightarrow\infty$,  hence we cannot use this equation to evaluate $\varphi_{TA}(0)$. Generally, we should use $\int_0^{\infty}S(\omega)d\omega=\pi \left\langle C_{TA}(t_m,0)\right\rangle$. Indeed,  for this case the ensemble average correlation function is $C(t,\tau=0)\propto t^{\alpha-1}$ as is given in \cite{Margolin04}. Then, following Eqs.~\eqref{eq:Eq9},\eqref{eq:ScaledCorr},\eqref{eq:CorrTA.EN}, we find $C_{\rm TA}(t,\tau=0)\propto t^{\alpha-1}$ as well. 

\begin{figure}
\centering
	\includegraphics[width=0.95\columnwidth]{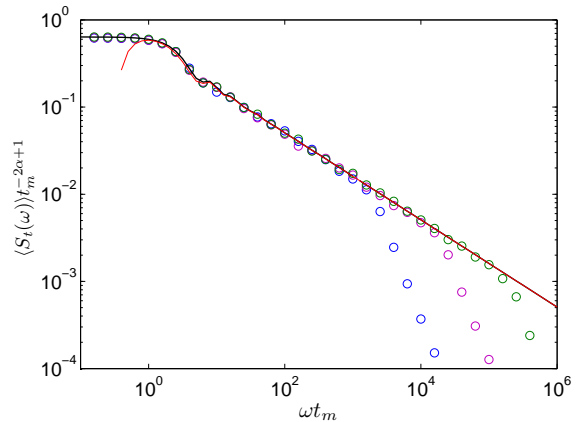}
	\caption{The simulation results (circles) for the $\left\langle S_{t_m}(\omega)\right\rangle$ in the blinking-quantum-dot model with finite-mean-``on'' times and the power-law-``off'' times . We use $\alpha=0.5$ at three different measurement times, $t_m=10^3$ (blue), $t_m=10^4$ (pink) and $t=10^5$ (green). 
The solid black line represents Eq.~(\ref{eq:PS_fin}) and red line is Eq.~\eqref{eq:PS_fin_osc}. As we increase measurement time the simulations approach theoretical prediction.  }
		\label{fig:PS.OnOff.alpha0.5}
\end{figure}

\section{Single File Diffusion}
\label{SFD}
The second example that we investigate is single-file diffusion. Single-file diffusion refers to the motion of particles in unidimensional systems, where the particles cannot pass each other, hence their ordering is preserved. We assume an infinite system, and we are interested in the displacement $x(t)$ of a tagged particle while all other particles playing the role of a bath \cite{Harris,Krapivsky,Hegde}. This kind of system can be used as a model for the motion of a single molecule in a crowed unidimensional environment such as a biological pore or channel \cite{Hodgkin,Macay}, and experimental studies of physical systems such as zeolites \cite{Hahn} and colloid particles in confined topology \cite{Wei} or optical tweezers \cite{Lutz}. %Although both examples present aging behavior, some of the aspects of the power spectrum are different between them.

We distinguish between two initial configurations of the bath: a thermal (equilibrium) initial condition (denoted as $(.)_{uni}$) and a non-thermal initial condition of equally spaced particles (labeled $(.)_{lat}$). 
%Initially, the particles are distributed according to a thermal (equilibrium) initial condition.
The free particle diffusion coefficient is $D$ and the average spacing between nearest particles is $a$. %The autocorrelation function of the tagged particle is different, not only in a prefactor, between the two cases \cite{Leibovich}, and in the limit $\tau\ll t$ we were found that
%\begin{eqnarray}
%C(t,\tau)_{lat}&\approx& %a\sqrt{\frac{D}{\pi}}\sqrt{t}\left(\sqrt{2}-\sqrt{\frac{\tau}{t}}\right) \\ \nonumber
%C(t,\tau)_{uni}&\approx& a\sqrt{\frac{D}{\pi}}\sqrt{t}\left(2-\sqrt{\frac{\tau}{t}}\right).
%\end{eqnarray} 
We note that the tagged particle is affected by the surrounding particles only at long times, i.e. $t\gg a^2/(2D)$. At shorter times $t\ll a^2/(2D)$ the tagged particle diffuses normally.

%\begin{figure}
%	\centering
%		\includegraphics[width=0.95\columnwidth,trim=0 150 0 20] {C:/MATLAB701/work/SIngleFileDiffusion/GeneralModelPhD.eps}
%	\label{fig:GeneralModelPhD}
%\end{figure}

\begin{figure}
	\centering
		\includegraphics[width=0.96\columnwidth]{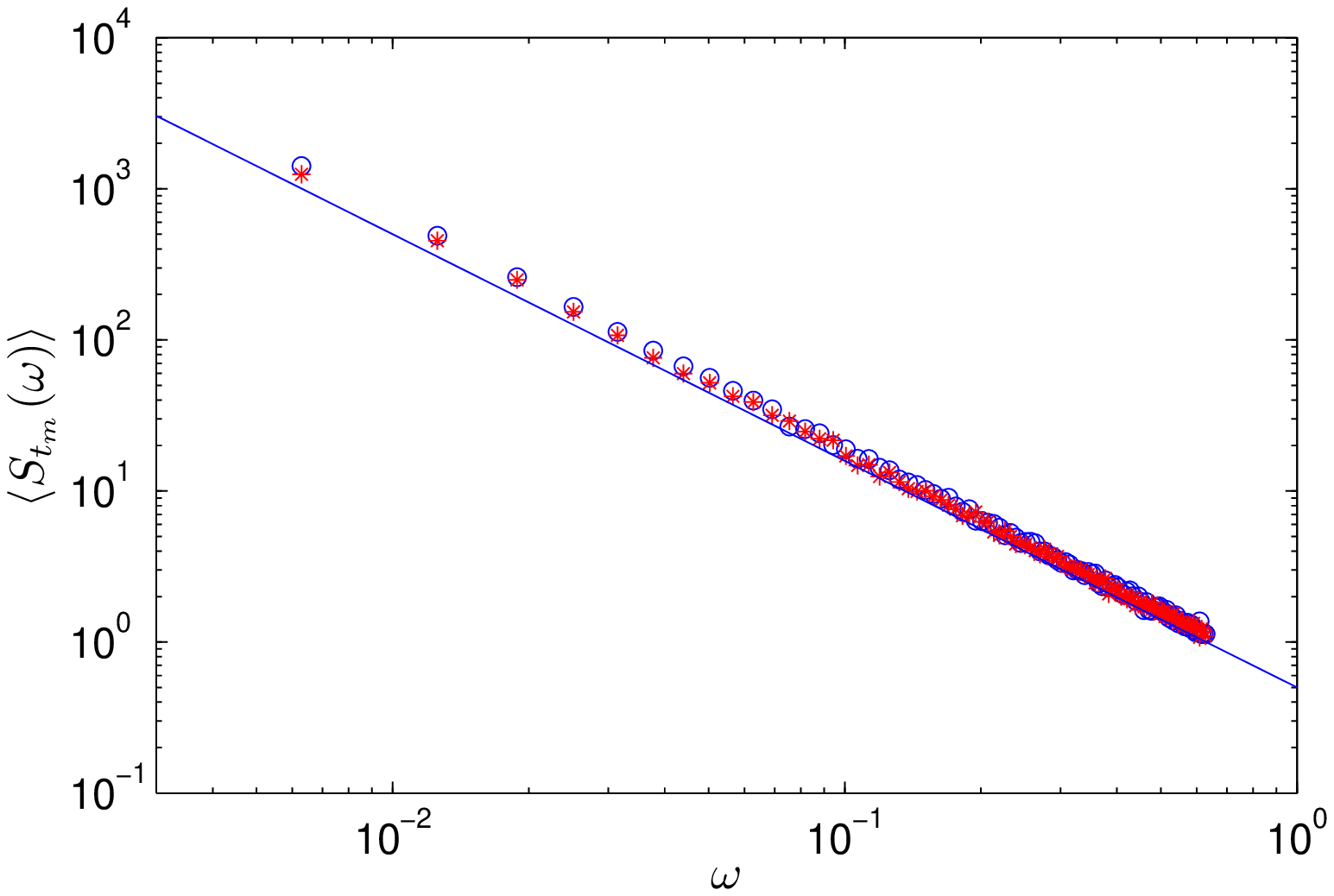}
		\includegraphics[width=\columnwidth]{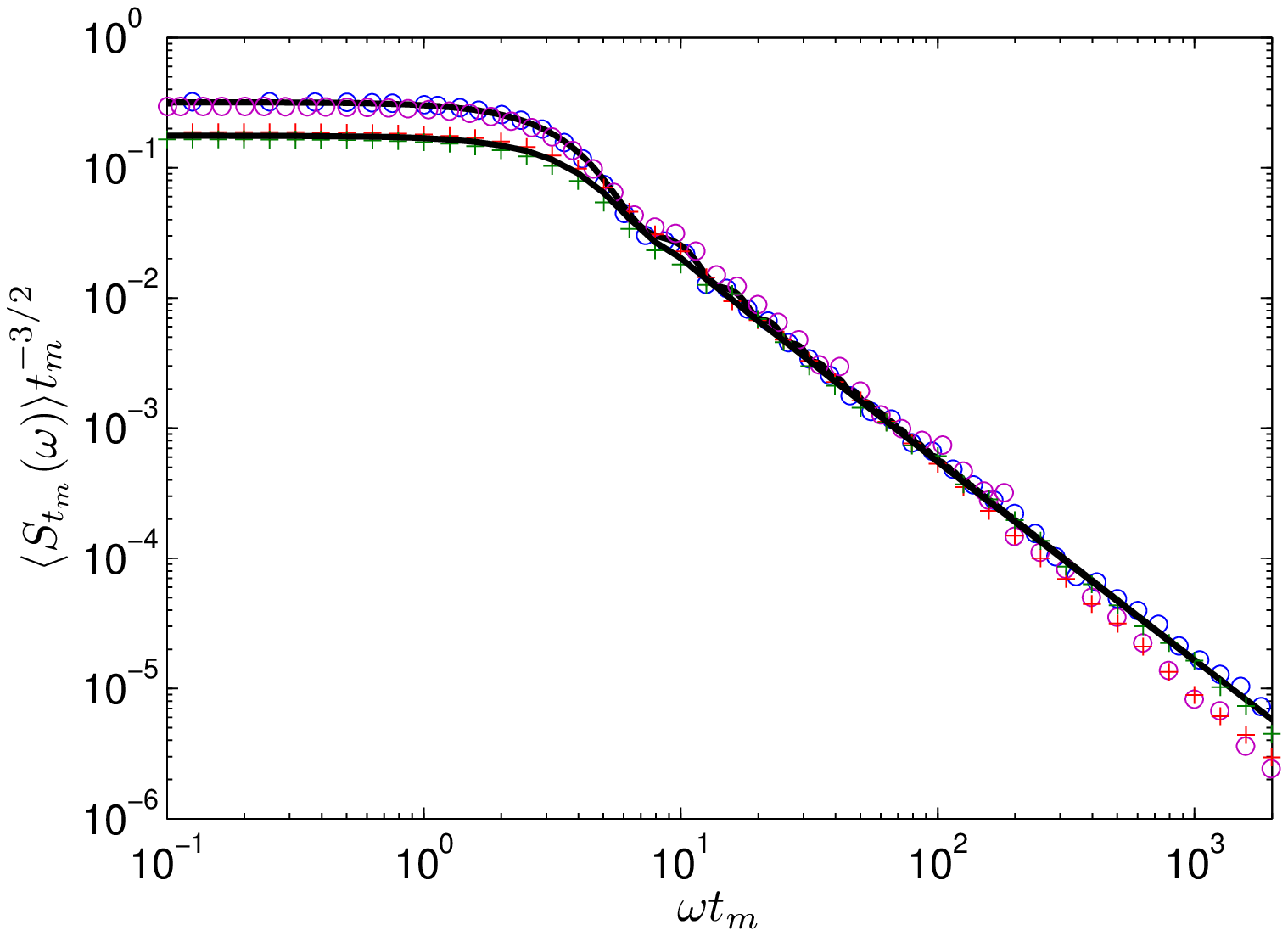}
		\caption{The simulation results for single-file diffusion for the two initial conditions; Upper panel: the spectrum $\langle S_{t_m}(\omega)\rangle$ versus the natural frequencies $\omega=2\pi n/t_m$, where $t_m=10^3$ and $n\in {\rm N}$ for two initial conditions, uniform (blue circles) and equidistance (red dots). For such presentation of the power spectrum, there is no distinction between two initial configuration. Lower panel: $\left\langle S_t(\omega)\right\rangle_{uni}$ (circles) at measurement times $t=10^3$ (blue) and $t=10^2$ (pink), and $\left\langle S_t(\omega)\right\rangle_{lat}$ (crosses) at measurement times $t=10^3$ (green) and $t=10^2$ (red). The solid black lines represent the analytic prediction Eq.~(\ref{eq:ICD.SFD}) and \eqref{eq:ICD.SFD.LAT}. The diffusion constant $D$ and the average distances between particles $a$ are taken to be $D=0.5$ and $a=1$. Deviations from theory at high frequencies are expected as explain in the text. They disappear as we take the measurement time to be long.
}
	\label{fig:NoiseSF.ICD}
\end{figure}

The correlation functions have been evaluated in \cite{Leibovich} and are given, for $t\gg a^2/(2D)$, by
\begin{eqnarray}
C(t,\tau)_{uni}&=&a\sqrt{\frac{D}{\pi}}\sqrt{t}\left(\sqrt{1+\frac{\tau}{t}}+1-\sqrt{\frac{\tau}{t}}\right), \label{eq:CorrSFD}\\
C(t,\tau)_{lat}&=&a\sqrt{\frac{D}{\pi}}\sqrt{t}\left(\sqrt{2+\frac{\tau}{t}}-\sqrt{\frac{\tau}{t}}\right). \nonumber
\end{eqnarray} 
By using Eq.~(\ref{eq:InfiniteCovarianceDensity}) we find that the spectrum for the equilibrium initial configuration is
\begin{eqnarray}
&& t_m^{-3/2}\sqrt{\frac{1}{Da^2}}\left\langle S_{t_m}(\omega)\right\rangle_{uni}=
\label{eq:ICD.SFD} \\&& =\frac{2+\cos (\tilde{\omega})}{\sqrt{\pi}\tilde{\omega}^{2}}   -\frac{1+2\cos(\tilde{\omega})}{\sqrt{2}\tilde{\omega}^{5/2}}{\cal{C}}\left(\sqrt{\frac{2\tilde{\omega}}{\pi}}\right) \nonumber \\  &&+\frac{\sqrt{2}}{\tilde{\omega}^{5/2}}{\cal{S}}\left(\sqrt{\frac{2\tilde{\omega}}{\pi}}\right)\left[-\tilde{\omega}+\sin(\tilde{\omega})\right], \nonumber
\end{eqnarray} 
where the Fresnel functions ${\cal{C}}(u)$ and ${\cal{S}}(u)$ are defined as 
\begin{eqnarray}
{\cal{C}}(u)=\int_0^u \cos(\pi t^2/2)dt \nonumber \\
{\cal{S}}(u)=\int_0^u \sin(\pi t^2/2)dt.
\end{eqnarray}
For the lattice initial condition we obtain,
\begin{eqnarray}
t_m^{-3/2}\sqrt{\frac{1}{Da^2}}\left\langle S_{t_m}(\omega)\right\rangle_{lat} =\frac{\sqrt{2}+  \cos (\tilde{\omega})}{\sqrt{\pi}\tilde{\omega}^{2}} \label{eq:ICD.SFD.LAT}\\   -\frac{\cos(2\tilde{\omega})}{\sqrt{2}\tilde{\omega}^{5/2}}\left[{\cal C}\left(\sqrt{\frac{4\tilde{\omega}}{\pi}}\right)-{\cal{C}}\left(\sqrt{
\frac{2\tilde{\omega}}{\pi}}\right)\right]  \nonumber \\  +\frac{\sin(2\tilde{\omega})}{\sqrt{2}\tilde{\omega}^{5/2}}\left[{\cal{S}}\left(\sqrt{\frac{2\tilde{\omega}}{\pi}}\right)-{\cal{S}}\left(\sqrt{\frac{4\tilde{\omega}}{\pi}}\right)\right] \nonumber \\
+\frac{\sqrt{2}}{\tilde{\omega}^{3/2}}\left[\tilde{\omega}{\cal S}\left(\sqrt{\frac{2\tilde{\omega}}{\pi}}\right)-{\cal C}\left(\sqrt{\frac{2\tilde{\omega}}{\pi}}\right)\right]. \nonumber
\end{eqnarray} 
As we see in Fig.~\ref{fig:NoiseSF.ICD}, these results are confirmed by simulations. The simulation method is described in Ref.~\cite{Leibovich}. As in the blinking quantum dot model we have assumed a scaling form of the correlation function Eq.~\eqref{eq:CorrSFD}, which works in the limit of large time. Information on the correlation function for short times is needed to estimate the very high frequency limit of the spectrum. Hence the deviations at high frequencies in Fig.~\ref{fig:NoiseSF.ICD} are expected. As the measurement time is increased, the spectrum plotted as a function of $\tilde{\omega}$ perfectly approaches the predictions of
our theory (see also the following example and Fig.~\ref{fig:PS_log_D_055_D_05_D_045}).

By using Eqs.~(\ref{eq:ScaledCorr1},\ref{eq:1FNoise}), we further find that the power spectrum corresponding to the random displacement $x(t)$ reads
\begin{equation}
\left\langle S_{t_m}(\omega)\right\rangle_{\omega t_m =2\pi n}^{uni}=\left\langle S_{t_m}(\omega)\right\rangle_{\omega t_m =2\pi n}^{lat}=\sqrt{\frac{a^2D}{2}}{\omega^{-3/2}},
\label{eq:PS_SFD}
\end{equation}
In the limit $\omega t_m=2\pi n \gg 1$, the power spectrum  seems to be time independent. However, we note that the spectrum $\langle S_{t_m}(\omega)\rangle$ remains time dependent. Hence, for every finite time $t_m$, the total power is finite, because of the low-frequency cutoff at $\omega \sim 1/t_m$, see Sect. V.C.
Moreover, measurements of the spectrum  $\langle S_{t_m}(\omega)\rangle$  made without knowledge of initial conditions would be consistent within the range of frequencies allowed by the limited observation time. In other words, the spectrum in the high frequencies limit, $\omega \gg 1/t_m$, is not affected by the initial condition, although the process is nonstationary.

\subsection{Critical Exponents for Single File Diffusion}
From Eq.~(\ref{eq:PS_SFD}) we observe that $\beta=3/2$ and $z=0$ for both cases.
Calculating $\left\langle S_{t_m}(0)\right\rangle$ by substituting $\omega=0$ in Eq.~(\ref{eq:GenerlizedWK}), yields 
\begin{eqnarray}
\left\langle S_{t_m}(0)\right\rangle_{uni}&=&\sqrt{\frac{Da^2}{\pi}}\frac{2}{5} t_m^{3/2}\nonumber \\
\left\langle S_{t_m}(0)\right\rangle_{lat}&=&\sqrt{\frac{Da^2}{\pi}}\frac{8}{15}(\sqrt{2}-1) t_m^{3/2}.
\end{eqnarray} 
We may then conclude that $\mu=3/2$ for both initial conditions.  The low-frequency cutoff is furthermore found to be 
\begin{eqnarray}
\omega_{\rm min}^{uni} &=& \left(\frac{25\pi}{8}\right)^{1/3}t_m^{-1} \nonumber \\
\omega_{\rm min}^{lat} &=& \left(\frac{15\sqrt{\pi}}{8(2-\sqrt{2})}\right)^{2/3}t_m^{-1}
\end{eqnarray}
and consequently $\eta=1$. The total measured power diverges here with time, i.e. 
\begin{equation}
\int_{\omega_{\rm min}}^{\omega_{\rm max}}\left\langle S_{t_m}(\omega)\right\rangle{\rm d}\omega\sim t_m^{1/2},
\end{equation}
for both initial configurations. As expected, the total power diverges since the displacement $x(t)$ is unbounded, even though for every finite time the total power is finite.

%\begin{figure}
    %\centering
    %\begin{tikzpicture}
        %\node[anchor=south west,inner sep=0] (image) at (0,0) {\includegraphics[width=0.95\columnwidth]{NoiseSF1.eps}};
        %\begin{scope}[x={(image.south east)},y={(image.north west)}]
            %\node[anchor=south west,inner sep=0] (image) at (0.2,0.2) {\includegraphics[width=0.35\columnwidth]{NoiseSF1.eps}};
        %\end{scope}
    %\end{tikzpicture}
    %\caption{...}
%\end{figure}

\section{Brownian Motion in A Logarithmic Potential}
\label{LogPotential}

\begin{figure}
	\centering	
	\includegraphics[width=0.95\columnwidth]{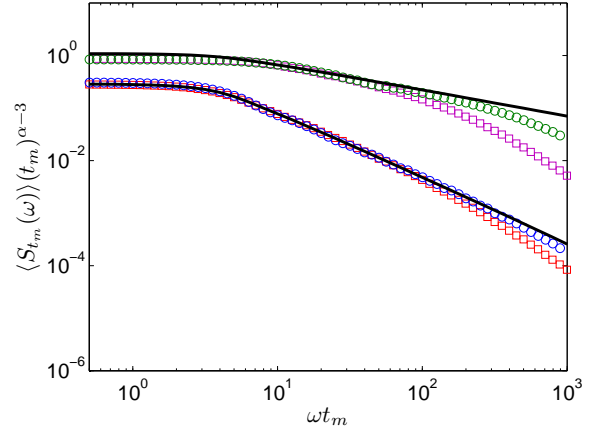}
		\caption{The simulation results for the diffusion in logarithmic potential when $D=0.4$ and measurement times $t_m=500$ (red squares) and $t_m=2000$ (blue circles) and when $D=0.25$ and measurement times $t_m=500$ (pink squares) and $t_m=2000$ (green circles). The black solid lines represent Eq.~(\ref{eq:LogSpectrum}). }
	\label{fig:PS_log_D_055_D_05_D_045}
\end{figure}

The third model that we consider is a Langevin equation with a logarithmic binding potential. Such a case is related, for example, to optical lattices, where $x$ is the momentum \cite{Marksteiner,Kessler}.  It further describes the denaturing of DNA \cite{Fogedby}, Manning condensation  on a polymer \cite{Manning}  or vortex dynamics \cite{Bray}. Such a case is interesting since it is ``weakly'' bound, and hence may exhibit different properties than free Brownian motion or Langevin dynamics in a harmonic potential \cite{Dechant,Dechant2012}. 

The Langevin equation describes the temporal evolution of the variable $x$ in a logarithmic binding potential for large $x$, 
\begin{equation}
\xi\frac{dx}{dt}+\frac{dU(x)}{dx}=\eta(t),
\label{eq:Langevin}
\end{equation}
where $\xi$ is a friction coefficient (we took a unit mass) and $\eta(t)$ is the white noise with zero mean,  satisfying the fluctuation-dissipation relation.  %Such a case is related, for example, to optical lattices \cite{Dechant}, where $p$ is the momentum.
We assume that the potential is of the form  
\begin{equation}
U(x)=\frac{1}{2}U_0\log(1+x^2),
\end{equation}
where the constant $U_0$ is related to the depth of the potential.   
In equilibrium, the PDF of $x$ is $P_{eq}(x)\sim\exp(-U(x)/k_BT)$ due to the Boltzmann theorem, therefore we find $P_{eq}(x)\sim(1+x^2)^{-U_0/2k_BT}$, i.e. $x$ has infinite variance when $1<U_0/k_BT<3$. 
We introduce the parameter $\alpha$ 
\begin{equation}
\alpha=\frac{U_0}{2k_BT}+\frac{1}{2}
\end{equation}
and define a diffusion constant through the fluctuation-dissipation relation, $D=k_BT/\xi$. 

The correlation function for the case where $\alpha>1$ is given in \cite{Dechant2012}, where both $t$ and $\tau$ are assumed to be large 
\begin{eqnarray}
&&C(t,\tau)\approx \frac{\sqrt{\pi}(4Dt)^{2-\alpha}}{Z\Gamma(\alpha)\Gamma(1+\alpha)}  \\ \nonumber
&& x^{2-\alpha}\int_0^{\infty}{\rm d}y e^{-y^2}y^2{_1F_1}\left(\frac{3}{2},\alpha+1,y^2\right)\Gamma(\alpha,y^2x),
\end{eqnarray}
where $x=\tau/t$.
Then, by using Eqs.~\eqref{eq:GenerlizedWK} and \eqref{eq:CorrTA.EN}, we find the spectrum
\begin{eqnarray}
\left\langle S_{t_m}(\omega)\right\rangle \label{eq:LogSpectrum} =&&t_m^{3-\alpha}\frac{2\sqrt{\pi}(4D)^{2-\alpha}}{Z\Gamma(\alpha)\Gamma(\alpha+1)} \times \\ \nonumber &&\int_0^1 {\rm d}x x^{3-\alpha}\cos(\tilde{\omega} x)\int_{\frac{x}{1-x}}^{\infty}{\rm d}y y^{-2}\times \\
&&\int_0^{\infty}{\rm d}z e^{-z^2}y^2{_1F_1}\left(\frac{3}{2},\alpha+1,z^2\right)\Gamma(\alpha,z^2y) \nonumber.
\end{eqnarray}
This is computed numerically by first evaluating the integral over
$y$ explicitly and then evaluated the remaining two integrals numerically by using Mathematica where the upper boundary for the $z$-integral is taken to be $10^5$ instead of $\infty$. The numerical results for $\alpha=1.75$ ($D=0.4$) and $\alpha=2.5$ ($D=0.25$) present good agreement with Langevin simulation results for not too large frequencies (see Fig.~\ref{fig:PS_log_D_055_D_05_D_045}). Initially, the particle is situated at the origin $x(t=0)=0$. Therefore, at short time the particle is not affected by the logarithmic tail of the potential. We thus expect a deviation from theory at high frequencies. 

We obtain for $1<\alpha<2$ in the limit $x \ll 1 $  
\begin{eqnarray}
&&C(t,\tau)\approx \frac{\sqrt{\pi}(4Dt)^{2-\alpha}}{Z\Gamma(\alpha)\Gamma(1+\alpha)}  \\ \nonumber
&&\left[\frac{\Gamma(\alpha+1)}{\sqrt{\pi}(2-\alpha)}+\frac{\sqrt{\pi}\Gamma(\alpha-2)\Gamma(\alpha+1)\Gamma(\alpha)}{4\Gamma^2(\alpha-\frac{1}{2})}x^{2-\alpha}\right],
\end{eqnarray}
where $Z=\sqrt{\pi}\Gamma(\alpha-1)/\Gamma(\alpha-1/2)$ is a normalization constant. 
For $2<\alpha<3$ the correlation function in the limit $\tau\ll t$ is stationary \cite{Dechant2012}
\begin{equation}
C(t,\tau)\approx \frac{\pi\Gamma(\alpha-2)(4D\tau)^{2-\alpha}}{4Z\Gamma^2(\alpha-\frac{1}{2})}.
\end{equation}
The corresponding power spectrum in discrete frequencies $\omega t_m=2\pi n \gg 1$ for both cases, $1<\alpha<2$ and $2<\alpha<3$, is moreover,
\begin{eqnarray}
&&\langle S_{t_m}(\omega)\rangle_{\omega t_m=2\pi n}= \label{eq:PS_log}\\ \nonumber &&2\sin\left(\frac{\pi\alpha}{2}\right)\Gamma(3-\alpha)\left(-\frac{\pi\Gamma(\alpha-2)(4D)^{2-\alpha}}{4Z\Gamma^2(\alpha-1/2)}\right)\omega^{\alpha-3}.
\end{eqnarray}
%\begin{eqnarray}
%&&C(t,\tau)\approx \frac{\sqrt{\pi}(4Dt)^{2-\alpha}}{Z\Gamma(\alpha)\Gamma(1+\alpha)}  \\ \nonumber
%&&\left[\frac{\Gamma(\alpha+1)}{\sqrt{\pi}(2-\alpha)}-\left(I_1-\frac{\Gamma(\alpha+1)\Gamma(\alpha)}{\sqrt{\pi}(2-\alpha)}\right)x^{2-\alpha}\right],
%\end{eqnarray}
%where $Z=\sqrt{\pi}\Gamma(\alpha-1)/\Gamma(\alpha-1/2)$ is a normalization constant, and $I_1=\int_0^1 {\rm d}y y^2 \exp(-y^2)M(3/2,\alpha+1,y^2)\Gamma(\alpha)$. 
%The corresponding power spectrum is
%\begin{eqnarray}
%&&\left\langle S_{t_m}(\omega)\right\rangle_{\omega t_m \gg 1}\approx 2\sin\left(\frac{\pi\alpha}{2}\right)\Gamma(3-\alpha) \nonumber\\ &&
%\frac{\sqrt{\pi}(4D)^{2-\alpha}}{Z\Gamma(\alpha)\Gamma(1+\alpha)}\left(I_1-\frac{\Gamma(\alpha+1)\Gamma(\alpha)}{\sqrt{\pi}(2-\alpha)}\right)\omega^{\alpha-3}.
%\label{eq:PS_log}
%\end{eqnarray}
We conclude that for both cases, $1<\alpha<2$ and $2<\alpha<3$, the critical exponents are: $\beta=3-\alpha$, $z=0$, $\eta=1$ and $\mu=3-\alpha$.
The total measured power, for both cases, is
\begin{equation}
\int_{\omega_{\rm min}}^{\omega_{\rm max}}\left\langle S_{t_m}(\omega)\right\rangle{\rm d}\omega\propto const. +t_m^{2-\alpha},
\end{equation}
where $\omega_{\rm min}\sim 1/t_m$ and $\omega_{\rm max}$ is time independent. When $1<\alpha<2$ the total power increases with the time $t_m$, i.e. it diverges when $t_m\rightarrow\infty$, since this case corresponds high temperature or shallow potential ($1<U_0/k_BT<3$) and thus the particle exhibits subdiffusion. When $2<\alpha<3$ (i.e. $3<U_0/k_BT<5$) we find that the total measured power in the frequencies range $(\omega_{\rm min},\omega_{\rm max})$ converges to a constant, i.e. effectively the particle is bounded. Therefore, its critical exponents are $\eta=2-\alpha$ for $1<\alpha<2$ and $\eta=0$ for $2<\alpha<3$.

\section{Summary and Discussion}
We have extended the \WK to nonstationary spectra by deriving two general relations between time- and ensemble- averaged correlation functions and the aging power spectrum \cite{LeibovichPRL,DechantPRL}. We have moreover established the generic occurrence of $1/f$ noise for nonanalytic ensemble averaged correlation functions and derived the corresponding five critical exponents. We have evaluated these exponents for three models: blinking quantum dot, single file diffusion and diffusion in a logarithmic potential.  The nonstationary spectrum retains all the important properties of the stationary one, in particular its interpretation as a density of Fourier modes. 
%To summarize we derived a general relation between the nonstationary correlation function and the aging power spectrum. This relation generalizes the \WK to aging systems. Since the ensemble averaged correlation function is not identical to the time averaged correlation function, we have two formalism.Care must be taken when we extract information on the underlying correlation function from the sample spectrum; this inverse problem is discussed briefly in Appendix A. We illustrate the effect of time dependency of the power spectrum in a blinking quantum dot model, single-file diffusion and a Brownian particle in a logarithmic potential. 
%From these results we draw the two following conclusions;  

%(i)
The nonstationarity of the correlation function does not necessarily imply that the spectrum in the $1/f$ regime is time dependent. Indeed we have found that the $1/f$ spectrum of single file diffusion and
diffusion in log potential are time independent. This happens because in these models, $\Upsilon={\rm V}$. Namely
the property of the correlation function determines if the aging exponent $z$ is zero or not. Therefore, by measuring time-independent $1/f$ noise, one cannot conclude that the process is stationary. %This issue was already discussed in Ref. \cite{Keshner}, where the author illustrated it with an infinite RC transmission line. 
This is well known for Brownian motion where the underlying process is nonstationary and the power spectrum is of $f^{-2}$ type. One way to reveal the nonstationarity is to present the data as $S/(t_m^{\Upsilon+1})$ versus $\tilde{\omega}=\omega t_m$ and see if a scaling solution in found. Another way is to search for the oscillations, see Eq.~\eqref{eq:NonStationatyPS}.    

%(ii) Even though measuring the spectrum at non-natural frequencies, i.e. $2\pi n/t_m \neq \omega_n$, is not customary since a period larger then the measurement time cannot be detected, the spectrum at these frequencies captures some properties of the correlation function, i.e. it does have a physical meaning. First, the behavior of the correlation function in the limit $1-\tau/t_m\ll 1$ provides the oscillatory part of the spectrum.  Second, the spectrum at $\omega=0$ is related to the time average correlation function.%, i.e. $t_m^{\Upsilon+1}\varphi_{\rm TA}(x)/\pi$.

%We note that $S_{t_m}(\omega)$ is a random function. Hence, an open question remains what is its distribution? For the renewal process the answer is known \cite{Niemann}. May we find some properties of $S_{t_m}(\omega)$ beside its ensemble average (which is discussed above) for other models? 
In addition, we have shown how the power spectrum for the single file system, depends on the initial condition. This theme could be further investigated for example in KPZ models \cite{TakeuchiJSP}, or when the measurement of blinking dots does not start at the beginning of the process, i.e. the effect of a waiting time on the power spectrum is important \cite{Bouchaud,Niemann2016}.

\section*{ACKNOWLEDGMENT}
This work was supported by the Israel Science Foundation.

\appendix
\begin{appendices}
\renewcommand{\theequation}{{\thesection}\arabic{equation}}
\section{Relation to Fourier Modes}
\label{sec:Fourier}
Following \cite{Kubo} we consider a signal  $I(t)$ which is observed over an interval $(0,t_m)$. 
Expand $I(t)$ in a Fourier series as 
\begin{equation}
I(t)=\sum_{n=-\infty}^{\infty} a_n e^{\imath\omega_n t}
\end{equation}
where the natural frequencies are defined $\omega_n=2\pi n/t_m$, and the Fourier coefficients are 
\begin{equation}
a_n=\frac{1}{t_m}\int_0^{t_m}I(t)e^{-\imath\omega_n t}dt,
\end{equation}
then $I_{t_m}(\omega)=a_nt_m$ (see Sect. \ref{sec2} in the main text).
We define the spectrum for finite time as
\begin{equation}
\left\langle S_{t_m}(\omega)\right\rangle\equiv t_m\left\langle |a_n|^2\right\rangle,
\label{eq:KuboSpectrum}
\end{equation}
where the measurement time $t_m$ is assume to be long.

We consider the blinking-quantum-dot model with long waiting time. Thus the ensemble averaged signal is simply a constant, $\langle I(t) \rangle=1/2$ (when starting in the ``on'' state this is valid in the long time limit \cite{Margolin04}). The average Fourier coefficient then is
\begin{equation}
\langle a_n\rangle = \langle I\rangle \delta_{n0}.
\label{eq:MeanFourier}
\end{equation}

When a suitable filter is used one may select a large number of the Fourier modes related to frequencies lying in the interval $\Delta\omega$, and hence observe a smooth power spectral density;
\begin{equation}
S_{t_m}(\omega) \Delta\omega= \left(\sum_{\omega_n\in\Delta\omega} a_n\right)^2=\sum_{n}\left|a_n\right|^2+\sum_n\sum_{m\neq n} a_na_m^\ast,
\end{equation}
where the number of modes in an interval $\Delta\omega$ is $t_m\Delta\omega/(2\pi)$.
To recover Eq.~\eqref{eq:KuboSpectrum} one should prove that the Fourier coefficients $\left\{a_n\right\}$ are mutually independent, i.e. 
\begin{equation}
\langle a_n a_m^\ast\rangle=\left\langle a_n\right\rangle\left\langle a_m^\ast\right\rangle =0,
\end{equation}
where $n\neq m$ (a topic left for future work). The last equality is a direct outcome of Eq.~\eqref{eq:MeanFourier}

\label{sec:bounds}
\subsection{Zero-Frequency Contribution}
We briefly remind the reader, some basic properties of Wienerian processes, using an example. 
Consider a process $I(t)$ with a stationary correlation function
\begin{equation}
\langle I(t)I(t+\tau)\rangle=\left[\left\langle I^2\right\rangle-\left\langle I\right\rangle^2\right]e^{-\tau}+\left\langle I\right\rangle^2.
\end{equation}
Following \WK Eq.~\eqref{eq:WK} we obtain the spectrum;
\begin{equation}
\langle S(\omega) \rangle =2\pi \left\langle I\right\rangle^2\delta(\omega)+\left[\left\langle I^2\right\rangle-\left\langle I\right\rangle^2\right]\frac{2}{1+\omega^2}.
\end{equation}
In experimental situations the zero-frequency power is not usually reported in the total power estimation. Therefore we expect to observe
\begin{equation}
H^+=\int_{0^+}^{\infty}S(\omega)d\omega=\pi\left[\left\langle I^2\right\rangle-\left\langle I\right\rangle^2\right].
\label{eq:B3}
\end{equation} 
Including the zero point gives
\begin{equation}
H^-=\int_{0^-}^{\infty}S(\omega)d\omega=\pi\left[\left\langle I^2\right\rangle+\left\langle I\right\rangle^2\right].
\label{eq:B14}
\end{equation}

Now, we have two ways to estimate $\langle I^2 \rangle$ from the power spectrum; The first one, is to shift the stationary process $I(t)$ in such a way that $\langle I\rangle=0$. The second one is to use the total power of a stationary process following the \WK \cite{Kubo}
\begin{equation}
\int_0^{\infty}\left\langle S(\omega)\right\rangle d\omega = \frac{1}{2}\int_{-\infty}^{\infty}\left\langle S(\omega)\right\rangle d\omega =\pi C(0)=\pi \langle I^2 \rangle,
\label{eq:B5}
\end{equation}
where $\left\langle S(\omega)\right\rangle$ is an even function of frequency. %Eq.~\eqref{eq:B5} is equivalent to take into the account only half of the zero-frequency spectrum, i.e $(H^++H^-)/2$. For a nonstationary process we use \eqref{eq:Total}, where similarly we used a half of the zero-frequency spectrum.

\subsection{Continuous versus Natural-Frequencies Spectrum }
In its discrete form one may compute by Euler-Maclaurin formula
\begin{equation}
\int_0^{\infty}\left\langle S_{t_m}(\omega)\right\rangle d\omega \approx \sum_{n=1}^{\infty}\left\langle S_{t_m}(\omega_n)\right\rangle\Delta\omega+ \frac{1}{2}\Delta\omega \left\langle S_{t_m}(0)\right\rangle,
\end{equation}
where $\omega_n$ are the natural frequencies defined above. Using $S(0)\Delta\omega=|a_0|^2$ and $S(\omega_n)\Delta\omega=|a_n|^2$ we find the total power
\begin{equation}
\int_0^{\infty}\left\langle S(\omega)\right\rangle d\omega \approx 2\pi\left( \sum_{n=1}^{\infty}\left\langle |a_n|^2\right\rangle+\frac{1}{2}\left\langle |a_0|^2\right\rangle\right).
\end{equation}
Following Parseval's identity 
\begin{equation}
\sum_{n=-\infty}^\infty |a_n|^2=\frac{1}{t_m}\int_{0}^{t_m}|I(t)|^2dt
\label{eq:Parseval}
\end{equation}   
and symmetry, i.e. $a_n=a_{-n}$ we find
\begin{equation}
\sum_{n=1}^{\infty}\left\langle S(\omega_n)\right\rangle\Delta\omega+ \frac{1}{2}\Delta\omega \left\langle S(0)\right\rangle= \pi\frac{1}{t_m}\int_{0}^{t_m}\left\langle |I(t)|^2\right\rangle dt.
\end{equation} 
For a stationary process the mean-squared-displacement is time independent $\left\langle I^2\right\rangle=C(0)$ and $\langle |a_0|^2 \rangle=\langle I \rangle^2$. Further from ergodicity and \eqref{eq:Parseval}  $\sum |a_n|^2=\langle I^2\rangle$, therefore
\begin{equation}
\sum_{n=1}^{\infty}\left\langle S(\omega_n)\right\rangle\Delta\omega=\pi\left(\left\langle I^2\right\rangle-\langle I\rangle^2 \right)=\pi C(0)-\pi\langle{I} \rangle^2 ,
\end{equation}
i.e. the total spectrum measurement provides the variance of the signal.

In a nonstationary process, we obtain 
\begin{eqnarray}
\sum_{n=1}^{\infty}\left\langle S_{t_m}(\omega_n)\right\rangle\Delta\omega=\pi\langle{\overline{I_{t_m}^2}}\rangle-\pi\left\langle \left({\overline{I_{t_m}}}\right)^2\right\rangle\\ \nonumber 
=\pi t_m^{\Upsilon}\varphi_{\rm TA}(0)-\pi\left\langle\left( {\overline{I_{t_m}}}\right)^2\right\rangle,
\end{eqnarray}
where the last equality is base on the scaling assumption Eq.~\eqref{eq:Eq9}, i.e. $\left\langle \overline{I_{t_m}^2}\right\rangle =t_m^{\Upsilon}\varphi_{\rm TA}(0)$ and $\overline{(.)}$ is the time-average defined as
\begin{equation}
\left\langle |a_0|^2\right\rangle=\left\langle\left( {\overline{I_{t_m}}}\right)^2\right\rangle=\left\langle \left(\frac{1}{t_m}\int_{0}^{t_m}I(t)dt \right)^2\right\rangle.
\end{equation}
In this case
\begin{equation}
\sum_{n=-\infty}^{\infty} |a_n|^2 = \frac{1}{t_m}\int_{0}^{t_m}|I(t)|^2dt=t_m^{\Upsilon}\varphi_{\rm TA}(0)
\label{eq:A19}
\end{equation}
We conclude that the time-dependent spectrum in its discrete form conserve the basic properties one expects the power spectrum to fulfill.  

\subsection{Illustration in Blinking-Quantum-Dot Model}
\begin{figure}
	\centering
		\includegraphics[width=\columnwidth]{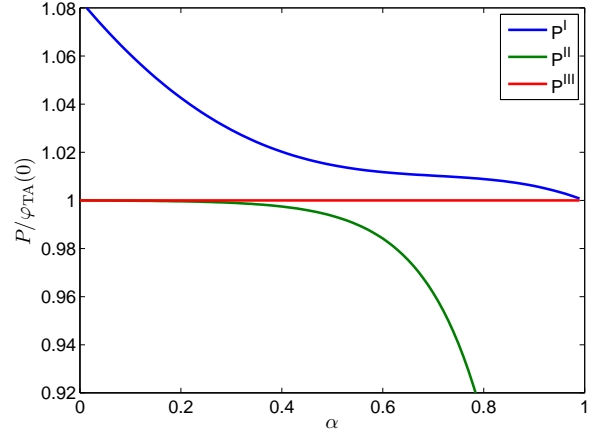}
		\caption{Comparison between the three estimation methods with the exact value of the total power in the blinking-quantum-dot model. }
	\label{fig:Math1}
\end{figure}

We use the blinking-quantum-dot model to to demonstrate numerically the estimation of the total power. Here we present three methods of estimation the correlation function from the power spectrum. We compare our results with the analytic results, see Fig.~\ref{fig:Math1}. Summing over the natural frequencies $\omega_n=2\pi n/t_m$ where $n\in{\mathbb{N}}$
\begin{eqnarray}
P_{\rm exact}&=& \frac{1}{2}\sum_{n=-\infty}^{\infty}\left\langle S_{t_m}(\omega_n)\right\rangle\Delta\omega= \label{eq:A20} \\  \nonumber
&=&\sum_{n=-\infty}^{\infty}\left\langle S_{t_m}(\omega_n)\right\rangle \frac{\pi}{t_m}=\pi/2,
\end{eqnarray}
where in this model $\phi_{\rm EA}(0)=\varphi_{\rm TA}(0)=\pi/2$ and $\Upsilon=0$ (see Eq.~\eqref{eq:CorrBQD}). 

The first method of evaluation of the total power is using the approximate spectrum Eq.~\eqref{eq:PS_inf} for $\left\langle S_{t_m}(\omega)\right\rangle$. It gives
\begin{equation}
P^I = 
\frac{\cos\left(\frac{\pi\alpha}{2}\right)(2\pi)^{\alpha-2}\zeta(2-\alpha)}{2\Gamma(1+\alpha)}+\frac{\pi}{4}(2-\alpha)
\end{equation}
where $\zeta(k)=\sum_{n=1}^{\infty}n^{-k}$ is the Riemann zeta function. The last term $\pi(2-\alpha)/4$ is related to the contribution from $n=0$. The deviation from the exact value Eq.~\eqref{eq:A20} may caused by the deviations of the approximate natural frequencies spectrum Eq.~\eqref{eq:PS_inf} from the exact spectrum Eq.~\eqref{eq:ICD.BQD} for small $n$. It means that the 1/f noise formula Eq.~\eqref{eq:PS_inf} is not sufficient for a precise estimate of the total power (see Fig. \ref{fig:Math1}).

As a second method for the estimation of the total power one may use use the exact expression for the spectrum Eq.~\eqref{eq:ICD.BQD}. 
The problem with this method is that there is no analytic expression for the infinite summation. To proceed we use a cutoff, $\omega_N=2\pi N/t_m$ where $N=10^3$,  for the large frequencies.
\begin{equation}
P^{II} = 
\sum_{n=1}^{10^3}\left[\frac{{\rm sinc}^2(\pi n)}{4}+\frac{\Im\left[M(1-\alpha;2;2\imath\pi n)\right]}{4\pi n}\right]+\frac{\pi(2-\alpha)}{4}.
\end{equation}
A large deviation from the exact value is observed when $\alpha$ larger then $\approx 0.7$ since when $\alpha$ approaches $1$ large frequencies spectrum contributes to the total power. 

The third method is a combination of the two previous ones;  
\begin{eqnarray}
P^{III} &=&\sum_{n=1}^{10^3}
\left[\frac{{\rm sinc}^2(\pi n)}{4}+\frac{\Im\left[M(1-\alpha;2;2\imath\pi n)\right]}{4\pi n}\right]  \\ 
&+& \frac{\cos\left(\frac{\pi\alpha}{2}\right)(2\pi)^{\alpha-2}}{2\Gamma(1+\alpha)}\sum_{n=10^3+1}^{\infty} n^{-2+\alpha}+\frac{\pi}{4}(2-\alpha) \nonumber.
\end{eqnarray}
Comparing this estimation to the exact result $P_{\rm exact}=\pi/2$ we find deviations of $0.005$ percent. Comparison between the three methods is given in Fig.~\ref{fig:Math1}. We use Mathematica for the numerical estimation of the summations. 

\section{On Eq.~(\ref{eq:Inverse})}
\label{sec:inverse}
We here show how to estimate the 
correlation function from the power spectrum.
Eq.~\eqref{eq:GenerlizedWK} in the discrete form is
\begin{eqnarray}
&&\left\langle S_{t_m}(\omega)\right\rangle\approx \nonumber\\ 
&&2t_m^{\Upsilon+1}\sum_{n=1}^{N-1}\left(1-n\Delta x\right)\varphi_{\rm TA}\left(n\Delta x\right)\cos(\omega t_m n\Delta x) \nonumber \\
&& + t_m^{\Upsilon+1}\varphi_{\rm TA}(0)\Delta x
\end{eqnarray} 
where we use the Euler-Maclaurin formula with the discrete variable  $x_n=n\Delta x$ and $\Delta x=1/N$ where $N$ is large.  Now, we multiply by $\cos(\omega t_m j \Delta x)$ and integrate over frequencies
\begin{eqnarray}
\int_0^{\pi/(t_m\Delta x)} d\omega \cos(\omega t_m n\Delta x)\left\langle S_{t_m}(\omega)\right\rangle\approx \\ \nonumber
\pi t_m^{\Upsilon}\left(1-n\Delta x\right)\varphi_{\rm TA}\left(n\Delta x\right).
\label{eq:A2}
\end{eqnarray} 
where $n\neq 0$. Therefore, using $\tilde{\omega}=\omega t_m$, we obtain
\begin{equation}
\frac{1}{\pi \left(1-x\right)}\int_0^{N \pi} d\tilde{\omega} \cos(\tilde{\omega} x )\left\langle S_{t_m}(\tilde{\omega})\right\rangle\approx \\ 
t_m^{\Upsilon+1}\varphi_{\rm TA}\left(x\right),
\end{equation}
Now, we would like to decrease the $x-$steps of the numeric integration for a certain measurement time $t_m$, or equivalently increasing $N$.  % Hence we choose $x=n/Nt_m$ where $N$ is a large integer. Thus, in the same fashion, we obtain
%\begin{equation}
%\frac{1}{\pi \left(1-x\right)}\int_0^{2\pi N} d\omega \cos(\omega x t_m)\left\langle S_{t_m}(\omega)\right\rangle\approx
%t_m^{\Upsilon}\varphi_{\rm TA}\left(x\right).
%\end{equation}
Where $N\rightarrow\infty$ we recover Eq.~\eqref{eq:Inverse} in the text. 

\begin{figure}
	\centering
		\includegraphics[width=0.95\columnwidth]{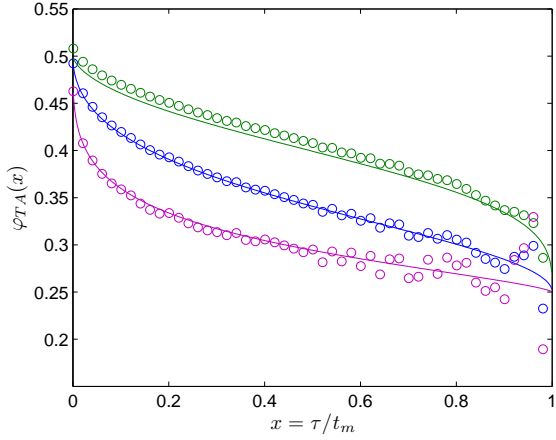}
		\caption{The simulation results (open circles) for Eq.~\eqref{eq:A2} in the blinking quantum dot process with heavy tailed PDF sojourn times Eq.~\eqref{eq:LongTailDistribution}. We examine three different exponents, $\alpha=0.3$ (green), $\alpha=0.5$ (blue) and $\alpha=0.8$ (pink). The solid lines represent the theory of time-average correlation function Eq.~\eqref{eq:CorrBQD}. The measurement time is $10^5$ and the ensemble average was taken over $10^4$ realizations.    }
	\label{fig:Eq11}
\end{figure}

In Fig. \ref{fig:Eq11} we show the estimation of the time averaged correlation function for a blinking quantum dot model where the sojourn times $\left\{\tau_i\right\}$ are distributed with the PDF Eq.~\eqref{eq:LongTailDistribution}. We first find the power spectrum  using the method presented  in App. C. We then apply Eq.~\eqref{eq:A2} to our `experimental' data, to find $\varphi_{\rm TA}(x)$ which is presented in Fig.~\ref{fig:Eq11}. 

The ensemble-average correlation function might be found by Eq.~\eqref{eq:Inv}, using the measured sample spectrum directly. Another method to obtain $\phi_{\rm EA}(x)$ is by taking the derivative of Eq.~\eqref{eq:CorrTA.EN}
\begin{eqnarray}
\phi_{\rm EA}(y)= 
&&\varphi_{\rm TA}\left(\frac{y}{1+y}\right)\left[(\Upsilon+1)(1+y)^{\Upsilon}-\Upsilon y(1+y)^{\Upsilon-1}\right]   \nonumber\\  
&&-\varphi'_{\rm TA}\left(\frac{y}{y+1}\right)y(1+y)^{\Upsilon}.
\end{eqnarray}
For $\Upsilon=0$, e.g. in the blinking-quantum-dot model, we use
\begin{equation}
\phi_{\rm EA}(y)= 
\varphi_{\rm TA}\left(\frac{y}{1+y}\right) 
-y\frac{d}{dy}\left[\varphi_{\rm TA}\left(\frac{y}{y+1}\right)\right].
\end{equation}
Thus, in principle, from estimation of $\varphi_{\rm TA}(x)$ using power spectrum one may obtain $\phi_{\rm EA}(x)$.

%For $x=0$ we find
%\begin{equation}
%\int_0^{2\pi} d\omega\left\langle S_{t_m}(\omega)\right\rangle\approx
%\end{equation}

\begin{widetext}

\section{The Power Spectrum for Blinking Quantum Dot Model}
\label{Exact}
%\subsection{Small Frequencies}
We substitute Eq.~(\ref{eq:CorrBQD}) in (\ref{eq:InfiniteCovarianceDensity}) and obtain
\begin{equation}
\left\langle S_{t_m}(\omega)\right\rangle/t_m=\underbrace{\int_0^1\frac{\tilde{\omega}x\sin(\tilde{\omega}x)+\cos(\tilde{\omega}x)-1}{\tilde{\omega}^2x^2}{\rm dx}}_{{\rm I}}- 
\frac{1}{2}\underbrace{\int_0^1\frac{\sin(\pi\alpha)}{\pi}B\left(x;1-\alpha,\alpha\right)
\frac{\tilde{\omega}x\sin(\tilde{\omega}x)+\cos(\tilde{\omega}x)-1}{\tilde{\omega}^2x^2}{\rm dx}}_{{\rm II}}.
\end{equation} 
The first term, I, contributes
\begin{equation}
{\rm I}=\int_0^1\frac{\tilde{\omega}x\sin(\tilde{\omega}x)+\cos(\tilde{\omega}x)-1}{\tilde{\omega}^2x^2}{\rm dx}=\frac{1}{2}{\rm sinc}^2\left(\frac{\tilde{\omega}}{2}\right).
\label{eq:B1}
\end{equation}
For the second term, II, we use integration by parts and obtain
\begin{equation}
{\rm II}=\frac{\sin(\pi\alpha)}{\pi}B(1;1-\alpha,\alpha)\frac{1-\cos(\tilde{\omega})}{\tilde{\omega}^2}-  \frac{\sin(\pi\alpha)}{\pi}\frac{1}{\tilde{\omega}^2}\int_0^1{\rm dx} \left[1-\cos(\tilde{\omega})\right]x^{-\alpha-1}(1-x)^{\alpha-1}.
\end{equation}
By definition the Kummer confluent function $M(a,b;z)$ for imaginary variable is
\begin{equation} 
\frac{\Gamma(b)}{\Gamma(a)\Gamma(b-a)}\int_0^1{\rm du}e^{\imath\tilde{\omega} u}u^{a-1}(1-u)^{b-a-1}\equiv M(a,b;\imath\tilde{\omega}).
\label{eq:B4}
\end{equation}
Taking the integration over $\tilde{\omega}$ in both sides of Eq/~\eqref{eq:B4}, with $a=1-\alpha$ and $b=1$, gives
\begin{equation}
\frac{1}{\Gamma(1-\alpha)\Gamma(\alpha)}\int_0^1{\rm d}x x^{-\alpha}(1-x)^{\alpha-1}\frac{e^{\imath\tilde{\omega}x}-1}{\imath x}=\int_0^{\tilde{\omega}}{\rm d}\tilde{\omega}_1M(1-\alpha,1,\imath\tilde{\omega}_1)=\tilde{\omega} M(1-\alpha,2;\imath\tilde{\omega}),
\end{equation}
and then taking the imaginary part
\begin{equation}
\frac{1}{\Gamma(1-\alpha)\Gamma(\alpha)}\int_0^1{\rm d}x x^{-\alpha-1}(1-x)^{\alpha-1}[\cos(\tilde{\omega}x)-1]=\Im\left[\tilde{\omega} M(1-\alpha,2;\imath\tilde{\omega})\right].
\end{equation}
Hence we conclude that 
\begin{equation}
{\rm II}=\frac{1}{2}{\rm sinc}^2\left(\frac{\tilde{\omega}}{2}\right)-\frac{1}{\tilde{\omega}}\Im\left[M(1-\alpha,2;\imath\tilde{\omega})\right].
\end{equation}
Evaluating I$-\frac{1}{2}$II in Eq.~\eqref{eq:B1} gives Eq.~(\ref{eq:ICD.BQD}) in the main text.

\end{widetext}

\section{Simulation Methods}
We use the sample power spectrum definition, i.e. $S_{t_m}(\omega)=|I_{t_m}(\omega)|^2/t_m$. In each system we generate the time series of the signal $I(t)$ and use discrete Fourier transform to find $I_{t_m}(\omega)$ and its complex conjugate $I_{t_m}(\omega)^*$. The simulation was done by Matlab standard fast Fourier transform (FFT) function. 
The power spectrum simulating in the renewal process may be faster by using the method below instead of using FFT function.

\subsection{Blinking Quantum Dot Simulation}
As was mentioned in the text, the process is defined with two states, $I_{off}=0$ and $I_{on}=1$, with random sojourn times in each state $\left\{\tau_i\right\}$. The system switches states alternately, ``off''$\leftrightarrow$``on'', every time $t_{n}=\sum_i\tau_i$. The random sojourn times $\tau_i$ are generated with $\tau=x^{-1/\alpha}$ where $x$ is random uniformly distributed in the interval $(0,1)$.
With that generation process we find
\begin{eqnarray}
\psi(\tau)=\alpha\tau^{-1-\alpha} &\ \ \ &\tau>1,
\end{eqnarray}
i.e. $\tau_0=1$, and its Laplace transform is
\begin{equation}
\psi(s)=\alpha E_{1+\alpha}(s),
\end{equation}
where $E_{1+\alpha}(s)\equiv\int_1^{\infty}t^{-1-\alpha}\exp(-st)dt$.
When $s\rightarrow 0$ we find
\begin{equation}
\psi(s)=1-\Gamma(1-\alpha)s^{\alpha}.
\end{equation}

The power spectrum for a single realization is defined as
\begin{equation}
t_mS_{t_m}(\omega)= 
\int_0^{t_m}I(t)\exp(-\imath\omega t){\rm d}t\int_0^{t_m}I(t)\exp(\imath\omega t){\rm d}t.
\end{equation} 
Since $I_{off}=0$ and $I_{on}=1$, as was mention above, we find
\begin{equation}
t_mS_{t_m}(\omega)=
\sum_{odds}\int_{t_i}^{t_{i+1}}\exp(-\imath\omega t){\rm d}t\sum_{odds}\int_{t_{j}}^{t_{j+1}}\exp(\imath\omega t){\rm d}t.
\end{equation} 
Calculating the integrals and rearranging the equation give
\begin{eqnarray}
t_mS_{t_m}(\omega)=&& \label{eq:BQDSimulation}\\ \nonumber
\frac{1}{\omega^2}\sum_{i,j}&&e^{-\imath\omega(t_{i+1}-t_{j+1})}+e^{-\imath\omega(t_i-t_j)} \\ \nonumber
&& -e^{-\imath\omega(t_i-t_{j+1})}-e^{-\imath\omega(t_{i+1}-t_{j})}.
\end{eqnarray} 
Using Equation (\ref{eq:BQDSimulation}) simplifies the simulations since finding the renewal times $t_i=\sum_{k=1}^j\tau_k$ is faster then find the entire sequence of $I(t)$ and use FFT. At last, we average over the realizations set.

\subsection{Single-File Diffusion Simulation Details}
In the single-file process we generate the signal $x(t)$ by using the method in \cite{Leibovich}. We used the diffusion coefficient $D=0.5$ and the average distance between nearest particles as $a=1$.

\subsection{Langevin Equation with Logarithmic Potential}
We generated $x(t)$ in processes which are modeled with Eq.~(\ref{eq:Langevin}),
with discretization of the Langevin equation. Namely for single realization we use
\begin{equation}
x(t+dt)=x(t)-\frac{x(t)}{1+x^2(t)}dt+\eta(dt),
\end{equation}
where the random variable $\eta(dt)$ is normally distributed with zero mean and variance $2Ddt$.
Notice that we used friction coefficient $\xi=1$ and external potential $U=\log(1+x^2)/2$. %We used diffusion coefficient $D=0.4$.  

%\section{The Scaling of $C(t,\tau)$ and $\left\langle C_{TA}(t,\tau)\right\rangle$.}
%Consider that the correlation function is scales as 
%\begin{equation}
%C(t,\tau)=t^{\Upsilon}\phi_{\rm EA}(\tau/t).
%\end{equation}
%Hence, the time averaged correlation function is
%\begin{equation}
%\left\langle C_{TA}(t,\tau)\right\rangle=\frac{1}{t-\tau}\int_0^{t-\tau}{\rm d}t_1t_1^{\Upsilon}\phi_{\rm EA}(\tau/t)
%\end{equation}
%We change the integration variable $x=\tau/t_1$ and find
%\begin{equation}
%\varphi_{\rm TA}(y)=\frac{y^{\Upsilon+1}}{1-y}\int_{\frac{y}{1-y}}^{\infty}{\rm d}x\frac{\phi_{EA}(x)}{x^{\Upsilon+2}}
%\end{equation}
%
%Now we assume that when $\tau\ll t$ the correlation function scales as 
%\begin{equation}
%C(t,\tau)\approx t^{\Upsilon}\left[A-B\left(\frac{\tau}{t}\right)^{N}\right].
%\label{eq:C4}
%\end{equation}
%Therefore, the correspond time averaged correlation function is
%\begin{eqnarray}
%\left\langle C_{TA}(t,\tau)\right\rangle&\approx&\frac{1}{t-\tau}\int_0^{t-\tau}{\rm d} t_1t_1^{\Upsilon}\left[A-B\left(\frac{\tau}{t_1}\right)^{N}\right] \nonumber\\  &\approx&\frac{At^{\Upsilon}}{\Upsilon+1}-\frac{B\tau^{N}t^{\Upsilon-N}}{\Upsilon-N+1}.
%\end{eqnarray}
%Hence, when $\tau\ll t$ we find similar scaling behavior as \eqref{eq:C4};
%\begin{equation}
%C_{TA}(t,\tau)\approx t^{\Upsilon}\left[\tilde{A}-\tilde{B}\left(\frac{\tau}{t}\right)^{N}\right],
%\end{equation}
%where $\tilde{A}=A/\left[\Upsilon+1\right]$ and $\tilde{B}=B/\left[\Upsilon-N+1\right]$.

\end{appendices}

\end{document}